\RequirePackage{fix-cm}
\documentclass[twocolumn,epjc3]{svjour3}
\smartqed  
\usepackage{graphicx}
\usepackage{amsmath}    
\usepackage{amssymb}    
\usepackage{latexsym}
\usepackage[colorlinks,citecolor=blue,urlcolor=blue,linkcolor=blue]{hyperref}
\usepackage[numbers,sort&compress]{natbib}
\journalname{Eur. Phys. J. C}

\begin{document}

\title{Probing the isotropy of cosmic acceleration using different supernova samples}

\author{Z. Q. Sun\thanksref{addr1}
        \and
        F. Y. Wang\thanksref{e1,addr1,addr2} 
}

\thankstext{e1}{e-mail: fayinwang@nju.edu.cn}


\institute{School of Astronomy and Space Science, Nanjing University, Nanjing 210093, China \label{addr1}
           \and
           Key Laboratory of Modern Astronomy and Astrophysics (Nanjing University), Ministry of Education, Nanjing 210093, China \label{addr2}
}

\maketitle

\begin{abstract}

Recent studies indicated that an anisotropic cosmic expansion may exist.
In this paper, we use three data sets of type Ia supernovae (SNe Ia) to
probe the isotropy of cosmic acceleration. For the Union2.1 data set,
the direction and magnitude of the dipole are $(l=309.3^{\circ} {}^{+
15.5^{\circ}}_{-15.7^{\circ}} ,\ b = -8.9^{\circ} {}^{ +
11.2^{\circ}}_{-9.8^{\circ}}  )$, and $\ A=(1.46 \pm 0.56) \times
10^{-3}$ from dipole fitting method. The hemisphere comparison results
are $\delta=0.20,l=352^{\circ},b=-9^{\circ}$. For the Constitution data
set, the results are $(l=67.0^{\circ}{}^{+
  66.5^{\circ}}_{-66.2^{\circ}},\ b=-0.6^{\circ}{}^{+
25.2^{\circ}}_{-26.3^{\circ}})$, and $\ A=(4.4 \pm 5.0) \times 10^{-4}$
for dipole fitting and $\delta= 0.56,l=141^{\circ},b=-11^{\circ}$ for
hemisphere comparison. For the JLA data set, no significant dipolar or
quadrupolar deviation is found. We find previous works using ($l, b, A$)
directly as fitting parameters may get improper results. We also explore
the effects of anisotropic distributions of coordinates and redshifts on
the results using Monte-Carlo simulations. We find that the anisotropic
distribution of coordinates can cause dipole directions and make dipole
magnitude larger. Anisotropic distribution of redshifts is found to have
no significant effect on dipole fitting results.  \end{abstract}


\section{Introduction}
\label{sec:intro}

Type Ia supernovae (SNe Ia) are ideal standard candles
\cite{Phillips1993}.  In 1998, the accelerating expansion of the
Universe was discovered using the luminosity-redshift relation of SNe Ia
\cite{Riess1998,Perlmutter1999}.  The cosmological principle assumes
that the Universe is homogeneous and isotropic at large scales.  Based
on the cosmological principle and numerous observational facts, the
standard $\Lambda$CDM model has been established.  It can be used to
explain various observations.

However, it is worthy to examine the validity of the standard
$\Lambda$CDM model
\cite{Kroupa2013,Kroupa2012,Perivolaropoulos2014,Koyama2015} and its
assumptions, namely the cosmological principle. Deviation from cosmic
isotropy with high statistical confidence level would lead to a major
paradigm shift. At present, the standard cosmology confronts some
challenges.  Observations on the large-scale structure of the Universe,
such as ``great cold spot'' on cosmic microwave background (CMB) sky map
\cite{Vielva2004}, alignment of lower multipoles in CMB power spectrum
\cite{DeOliveira-Costa2004,Tegmark2003}, alignment of polarization
directions of quasars in large scale \cite{Hutsemekers2005}, handedness
of spiral galaxies \cite{Longo2009}, and spatial variation of the fine
structure constant \cite{King2012,Mariano2012}, show that the Universe
may be anisotropic.

The isotropy of the cosmic acceleration has been widely tested using SNe
Ia. Generally, there are two different ways to study the possible
anisotropy from SNe Ia. The first one is directly fitting the data to a
specific anisotropic model (AM) \cite{Campanelli2011,
Li2013, Wang2018}. Many anisotropic cosmological
models have been proposed to match the observations, including the
Bianchi I type cosmological model
\cite{Campanelli2011,Aluri2013} and the Rinders-Finsler
cosmological model \cite{Chang2014}. The extended topological
quintessence model with a spherical inhomogeneous distribution for dark
energy density is also proposed \cite{Mariano2012}.

An alternative method is directly analysing the SNe Ia data in a
model-independent way \cite{Antoniou2010, Cai2012, Cai2013,
  Mariano2012, Zhao2013, Yang2013, Wang2014, Javanmardi2015,
  BeltranJimenez2015, Lin2016, Chang2018, Deng2018a, Deng2018b,
Sun2018}, which does not depend on the specific cosmological model. The
hemisphere comparison (HC) method and dipole fitting (DF) method are
usually used in literature. The hemisphere comparison method divides
samples into two hemispheres perpendicular to a polar axis, then fits
cosmological parameters using samples in each hemisphere independently
and compares their differences. The dipole fitting (DF) method assumes a
dipolar deviation on redshift-distance modulus relation, then derives
the dipole's direction and magnitude using statistic approaches.
Meanwhile, low-redshift SNe Ia are used to estimate the direction and
amplitude of the local bulk flow \cite{Bonvin2006,
  Schwarz2007, Gordon2008, Colin2011,
  Turnbull2012, Kalus2013, Appleby2014,
Huterer2015}. So far, no study has been able to rule out the
isotropy at more than 3$\sigma$. The gravitational wave as standard
siren has also been proposed to probe cosmic anisotropy
\cite{Cai2018}.

The directions and magnitudes of anisotropy from previous works are
shown in Table \ref{table:previous-works}. It's obvious that different
results are derived from different authors. In this paper, we compare
the DF fitting results of different SNe Ia samples and try to find the
reason for the differences. This paper is organized as follows. In
Section~\ref{sec:methods}, SNe Ia data sets and DF method are
introduced. The fitting results are shown in section~\ref{sec:results}.
We discuss the possible reasons for the differences in
Section~\ref{sec:discussions}. Finally, we summarize in
Section~\ref{sec:conclusions}.

\section{Data and Methods}
\label{sec:methods}

\subsection{data sets}

Large-scale systematic sky surveys on SNe Ia have been performed in the
past decades. These surveys, which cover a wide range of redshifts from
$z<0.1$ to $z \sim 1$, include Supernovae Legacy Survey
\cite[SNLS]{Astier2006,Sullivan2011}, Sloan Supernova
Survey \cite{Holtzman2008}, the Pan-STARRS survey
\cite{Rest2013,Scolnic2013,Tonry2012}, Harvard-Smithsonian Center for
Astrophysics survey  \cite[CfA]{Hicken2009}, the Carnegie Supernova
Project \cite[CSP]{Stritzinger2011,Folatelli2009,Contreras2009}, the
Lick Observatory Supernova Search \cite[LOSS]{Ganeshalingam2013}, the
Nearby Supernova Factory\cite[NSF]{Aldering2002}, etc. Thanks to these
sky surveys, a bunch of SNe Ia catalogs has been published, including
``SNLS'' \cite{Astier2006}, ``Union'' \cite{Rubin2009},
``Constitution'' \cite{Hicken2009}, ``SDSS'' \cite{Campbell2013},
``SNLS3'' \cite{Conley2011}, ``Union2.1'' \cite{Suzuki2012}, and
``Joint Light-curve Analysis(JLA)'' \cite{Betoule2014}.

\begin{figure*}
   \includegraphics[width=\textwidth]{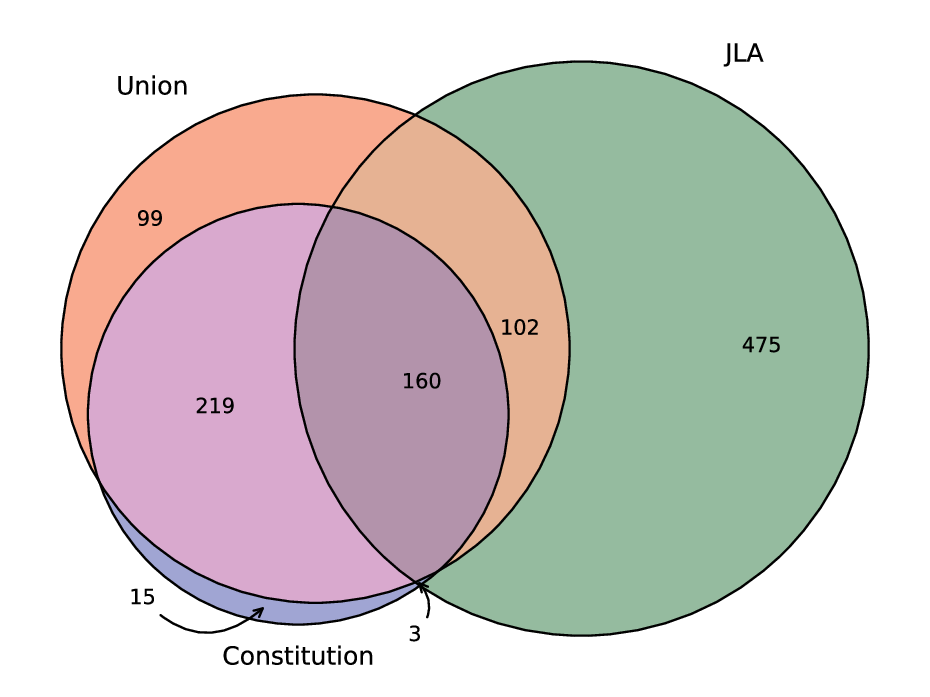}
   \caption {Venn diagram of the three SNe Ia catalogs. Overlapped data points are shown as labeled numbers.}\label{fig:Venn}
 \end{figure*}

In this paper, we used three SNe Ia catalogs in our analysis:
Union2.1, Constitution, and JLA. Union2.1 includes 580 SNe Ia
\cite{Suzuki2012}. The catalog covers samples with redshift range
$0.015 \leq z \leq 1.414$. Constitution catalog combines samples
from Union and CfA3, containing 397 SNe Ia with redshifts in the
range $0.015 \leq z \leq 1.55$ \cite{Hicken2009}. The coordinate
information of the Constitution catalog is adopted from the Open
Supernova Catalog in this paper \cite{Guillochon2016}. JLA catalog
includes several low-redshift samples ($z < 0.1)$, all three seasons
from the SDSS-II ($0.05 < z < 0.4$), and three years from SNLS ($0.2
< z < 1$). It includes 740 SNe Ia with high-quality light curves
\cite{Betoule2014}. It covers redshift range $0.01 \leq 1.30 $. The
three SNe Ia catalogs have some overlap in samples. The numbers
  of overlapped data points are shown in Figure \ref{fig:Venn}.

\subsection{Dipole fitting method}

Firstly, we briefly introduce the dipole fitting method.  For Union2.1
and Constitution data sets, the luminosity distance could be expanded by
Hubble series parameters: Hubble parameter $H$, deceleration parameter
$q$, jerk parameter $j$ and snap parameter $s$.  These parameters can be
expressed as functions of the scale factor $a$ and its derivatives,
\begin{equation}
\begin{aligned}
H& = \frac{\dot{a}}{a}&
q& = -\frac{1}{H^2}\frac{\ddot{a}}{a}\\
j& = \frac{1}{H^3}\frac{\dddot{a}}{a}&
s& = \frac{1}{H^4}\frac{\ddddot{a}}{a}.
\end{aligned}
\end{equation}
Taylor expansion of luminosity distance could be made in terms of
redshift and Hubble series parameters \cite{Visser2004}.  However, this
expansion diverges at $z>1$.  Thus, another parameter $y=z/(1+z)$ is
introduced to overcome this problem.  The luminosity distance can be
expanded as a function of $y$ \cite{Cattoen2007, Wang2009}
\begin{equation}
  \begin{aligned}
    d_\mathrm{L}(y) & = \frac{c}{H_0}[y-\frac{1}{2}(q_0-3)y^2
                      + \frac{1}{6}(11-5q_0-j_0)y^3 \\
                    & + \frac{1}{24}(50-7j_0-26q_0+10q_0j_0+21q_0^2-15q_0^3+s_0)y^4 \\
                    & + O(y^5)].
    \label{eq:dl_y}
  \end{aligned}
\end{equation}
The distance modulus is defined as
\begin{equation}
\mu_\mathrm{th} = 5 \log \frac{d_\mathrm{L}}{\mathrm{Mpc}} + 25.
    \label{eq:mu_th_dl}
\end{equation}
Then the $\chi^2$ can be calculated as
\begin{equation}
    \chi^2(H_0, q_0, j_0, s_0) = \sum_i \frac{(\mu_{\mathrm{obs},i} - \mu_{\mathrm{th},i})^2}{\sigma_i^2},
    \label{eq:chi2_Union}
\end{equation}
where $\mu_\mathrm{obs}$ and $\sigma_i$ are observational values of
distance moduli and their errors, respectively. The best-fitting values
of parameters could be obtained by minimizing $\chi^2(H_0, q_0, j_0,
s_0)$.

For the JLA sample, the observational values of distance moduli are not
directly given. Therefore, we use the values obtained in the
$\Lambda$CDM model to avoid fitting too many free parameters
simultaneously. The theoretical luminosity distance in $\Lambda$CDM
model can be expressed as
\begin{equation}
d_\mathrm{L}(z) = \frac{(1+z)c}{H_0} \int_{0}^{z}
\frac{\mathrm{d}z'}{\sqrt{\Omega_\mathrm{m}(1+z)^3+(1-\Omega_\mathrm{m})}}.
\end{equation}

Union2.1 and Constitution data sets already give $\mu_\mathrm{obs}$ as a
part of the released data. For JLA data set, $\mu_\mathrm{obs}$ can be
derived from light curve parameters of SN Ia from
\cite{Betoule2014}
\begin{equation}
\mu_\mathrm{obs} = m_\mathrm{B}^*-(M_\mathrm{B} - \alpha \times X_1 + \beta \times \mathcal{C}),
\end{equation}
where $m_\mathrm{B}^*$ is the observed peak magnitude in the rest-frame
of the B band, $X_1$ describes the time stretching of light-curve,
$\mathcal{C}$ describes the supernova color at maximum brightness and
$M_\mathrm{B}$ is the absolute B-band magnitude, which depends on the
host galaxy properties. $\alpha,\beta$ are nuisance parameters.
$M_\mathrm{B}$ can be fitted by a simple step function related to
$M_\mathrm{stellar}$ \cite{Johansson2013},
\begin{equation}
M_\mathrm{B} = \left \{
    \begin{array}{ll}
    M_\mathrm{B}^1            & M_\mathrm{stellar}       <   10^{10} M_{\odot} \\
    M_\mathrm{B}^1 + \Delta_M & M_\mathrm{stellar} \geqslant 10^{10} M_{\odot}. \\
    \end{array}
    \right.
\end{equation}
where $M_\mathrm{B}^1$ and $\Delta_M$ are nuisance parameters.
$\boldsymbol{C}(\alpha,\beta)$ is the total covariance matrix, which can
be obtained with JLA data. $\chi^2$ is defined as
\begin{equation}
    \chi^2_{JLA} (\alpha, \beta, M_\mathrm{B}^1, \Delta_M, \Omega_\mathrm{m}) = (\boldsymbol{\mu}_\mathrm{obs} -
\boldsymbol{\mu}_\mathrm{th})^{\dag} \boldsymbol{C}^{-1}
(\boldsymbol{\mu}_\mathrm{obs} - \boldsymbol{\mu}_\mathrm{th}).
    \label{eq:chi2_JLA}
\end{equation}

By minimizing $\chi^2_{JLA}$, all free parameters mentioned above can be
fitted. Our best-fitting results are consistent with those of
\cite{Betoule2014}.

To quantify the anisotropic deviations on luminosity distance, we
define the distance moduli with dipole $\boldsymbol{A}$ and monopole
$B$ as
\begin{equation}
\tilde{\mu}_\mathrm{th} = \mu_\mathrm{th}(1 - \boldsymbol{A} \cdot \hat{\boldsymbol{n}} + B),
\end{equation}
where $\tilde{\mu}_\mathrm{th}$ is the theoretical value of distance
modulus with dipolar direction dependence, and $\hat{\boldsymbol{n}}$ is
the unit vector pointing at the corresponding SN Ia. $\boldsymbol{A}
\cdot \hat{\boldsymbol{n}}$ represents the projected dipole magnitude in
the direction of the given SNe Ia sample. $\hat{\boldsymbol{n}}$ can be
represented in galactic coordinate as
\begin{equation}
\hat{\boldsymbol{n}} = \cos (b) \cos (l) \hat{\boldsymbol{i}} + \cos (b) \sin (l) \hat{\boldsymbol{j}} + \sin (b) \hat{\boldsymbol{k}}.
\end{equation}
Then the projection is
\begin{equation}
\boldsymbol{A} \cdot \hat{\boldsymbol{n}} = \cos (b) \cos (l)A_x + \cos (b) \sin (l) A_y + \sin (b) A_z.
\end{equation}

The best-fitting dipole and monopole parameters can be derived with the
following steps:
  \begin{enumerate}
    \item Substitute $\mu_\mathrm{th}$ with
      $\tilde{\mu}_\mathrm{th}$ in the expression of $\chi^2$ as is
      shown in equation~(\ref{eq:chi2_Union}) and
      equation~(\ref{eq:chi2_JLA}).
    \item Fit the dipole $(A_x,
      A_y, A_z)$ and the monopole component $B$ by minimizing
      $\chi^2$.
    \item Transform the fitted dipole from Cartesian coordinate
      $(A_x, A_y, A_z)$ back to spherical coordinate $(l, b, A)$.
  \end{enumerate}

Finally, we analyze the likelihood of the fitted parameters and the
significance of dipole magnitude utilizing Markov Chain Monte Carlo
(MCMC) sampling. To obtain the significance of dipole
anisotropy precisely, we use the Monte Carlo simulation method
\cite{Mariano2012,Yang2013,Wang2014}. To be specific, we construct a
type of synthetic samples based on original data sets by assuming that
theoretical values of distance moduli are ``real'' values. We refer to
these synthetic samples as ``isotropic'' samples. Applying MCMC sampling
on these samples, probability distributions of the fitted parameters can
be obtained. They are also used to probe the effect of anisotropic
factors, which we will discuss in section \ref{sec:discussions}.

\subsection{Quadrupole fitting}

 To examine whether a higher order of anisotropy exists, we fit the
 quadrupole along with the dipole in a similar manner as mentioned
 previously. We define the quadrupole $\mathbf{Q}$ in
\begin{equation}
\tilde{\mu}_\mathrm{th} = \mu_\mathrm{th}(1 - \mathbf{Q}^T \hat{\boldsymbol{n}} \mathbf{Q} - \boldsymbol{A} \cdot \hat{\boldsymbol{n}} + B),
\end{equation}
where $\mathbf{Q}$ is a symmetric, traceless $3\times3$ matrix
determined by 5 individual parameters.

\subsection{Hemisphere comparison method}

We applied hemisphere comparison method to the three samples. For a
randomly chosen axis, data points contained in each hemisphere is used
to fit $\Omega_\mathrm{m}$ individually, then the difference between
each hemisphere is shown as
\begin{equation}
  \label{eq1} \delta=\frac{\Delta
  \Omega_\mathrm{m0}}{\bar\Omega_\mathrm{m0}}=\frac{\Omega_\mathrm{m0,u}-\Omega_\mathrm{m0,d}}{(\Omega_\mathrm{m0,u}+\Omega_\mathrm{m0,d})/2},
\end{equation}
where the subscripts $u$ and $d$ represent the best parameter fitting
value in the `up' and `down' hemispheres, respectively.

\section{Results}

\subsection{Dipole fitting method}
\label{sec:results}

\begin{figure*}
 \includegraphics[width=\textwidth]{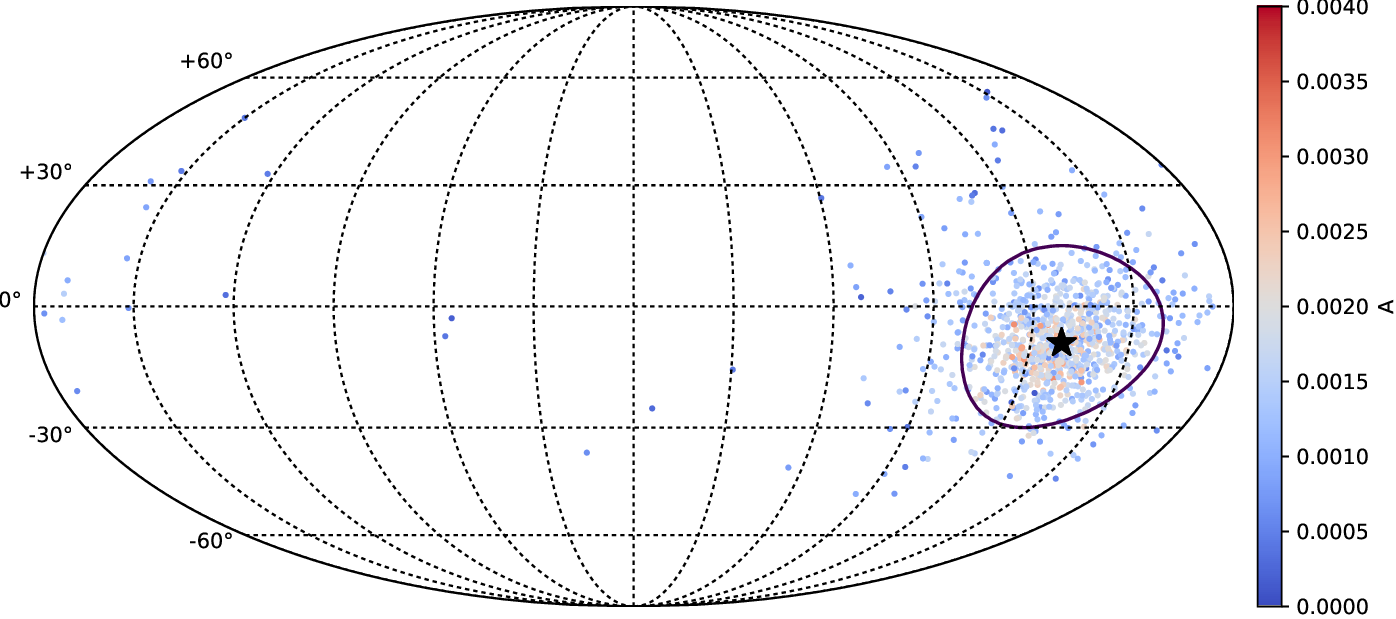}
 \caption{Best-fitting dipole direction (star) and 1$\sigma$ error
 range of the Union2.1 data set. Scatter points represent dipole
 fitting results of simulating samples.}\label{fig:UnionSamples}
 \end{figure*}

\begin{figure*}
 \includegraphics[width=\textwidth]{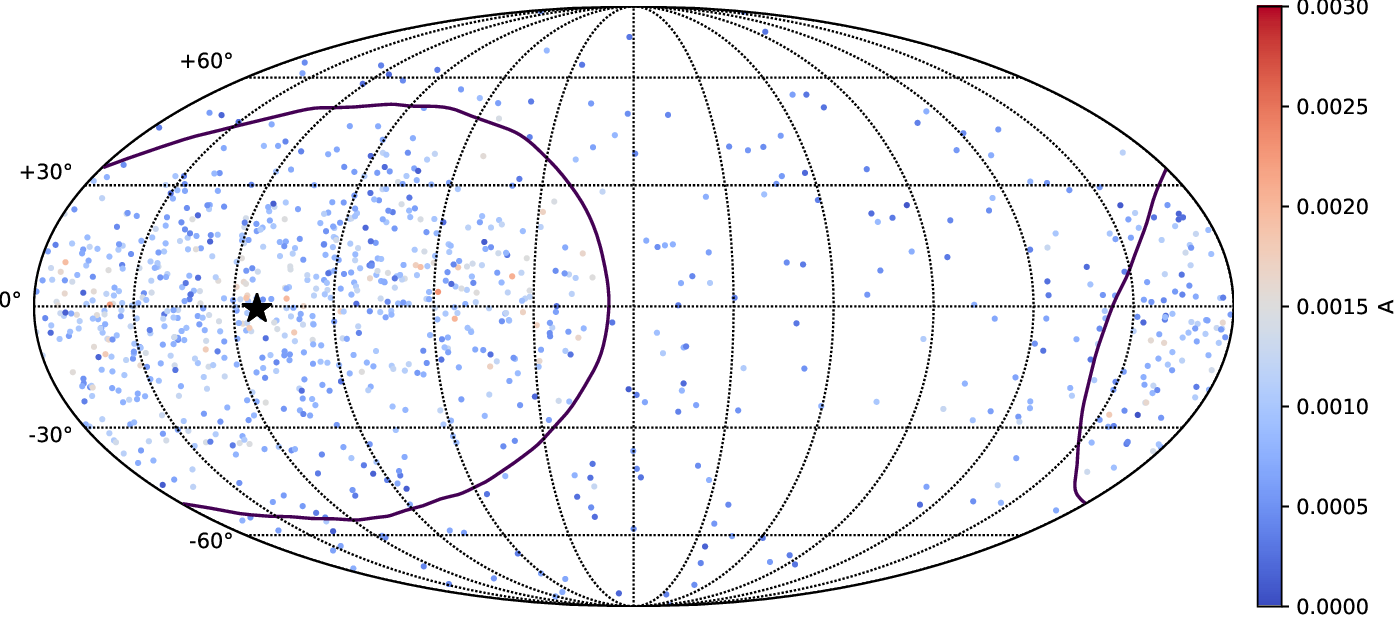}
 \caption{Similar to Figure \ref{fig:UnionSamples} for the
 Constitution data set. }\label{fig:ConstitutionSamples} \end{figure*}

\begin{figure*}
   \includegraphics[width=\textwidth]{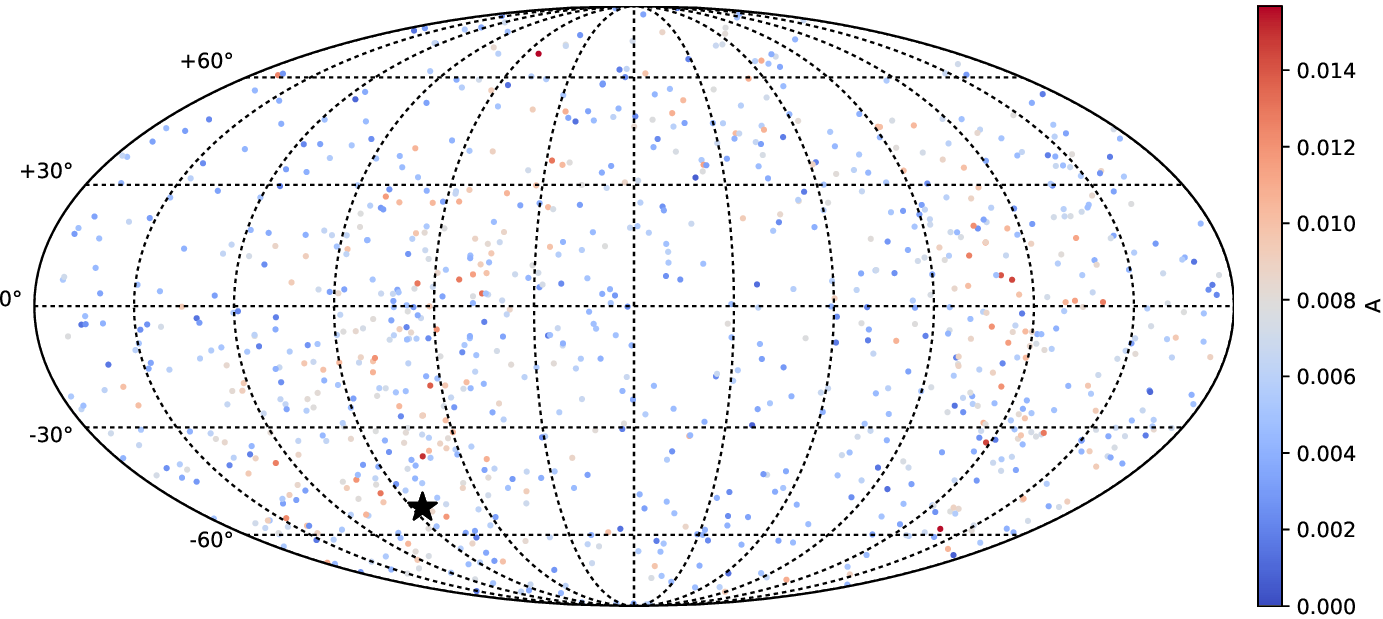}
   \caption{Best-fitting dipole direction (star) of the JLA data set.
   Scatter points represent dipole directions and magnitudes generated
 by MCMC sampling of original samples.} \label{fig:JLASamples}
 \end{figure*}

Fitting results are shown in Table \ref{table_results}. The confidence
level is defined as the probability
$\mathrm{P(|\boldsymbol{A}_{iso}|<|\boldsymbol{A}_{fit}|)}$, where
$\mathrm{|\boldsymbol{A}_{iso}|}$ is the dipole magnitude of an
arbitrary data set in ``isotropic'' samples, and
$\mathrm{|\boldsymbol{A}_{fit}|}$ is the best-fitting dipole magnitude.
Best-fitting dipole directions and 1$\sigma$ errors for Union2.1,
Constitution, and JLA data sets, along with dipole fitting results of
samples are plotted in Figures \ref{fig:UnionSamples},
\ref{fig:ConstitutionSamples} and \ref{fig:JLASamples}, respectively.

\begin{figure*} \includegraphics[width=\textwidth]{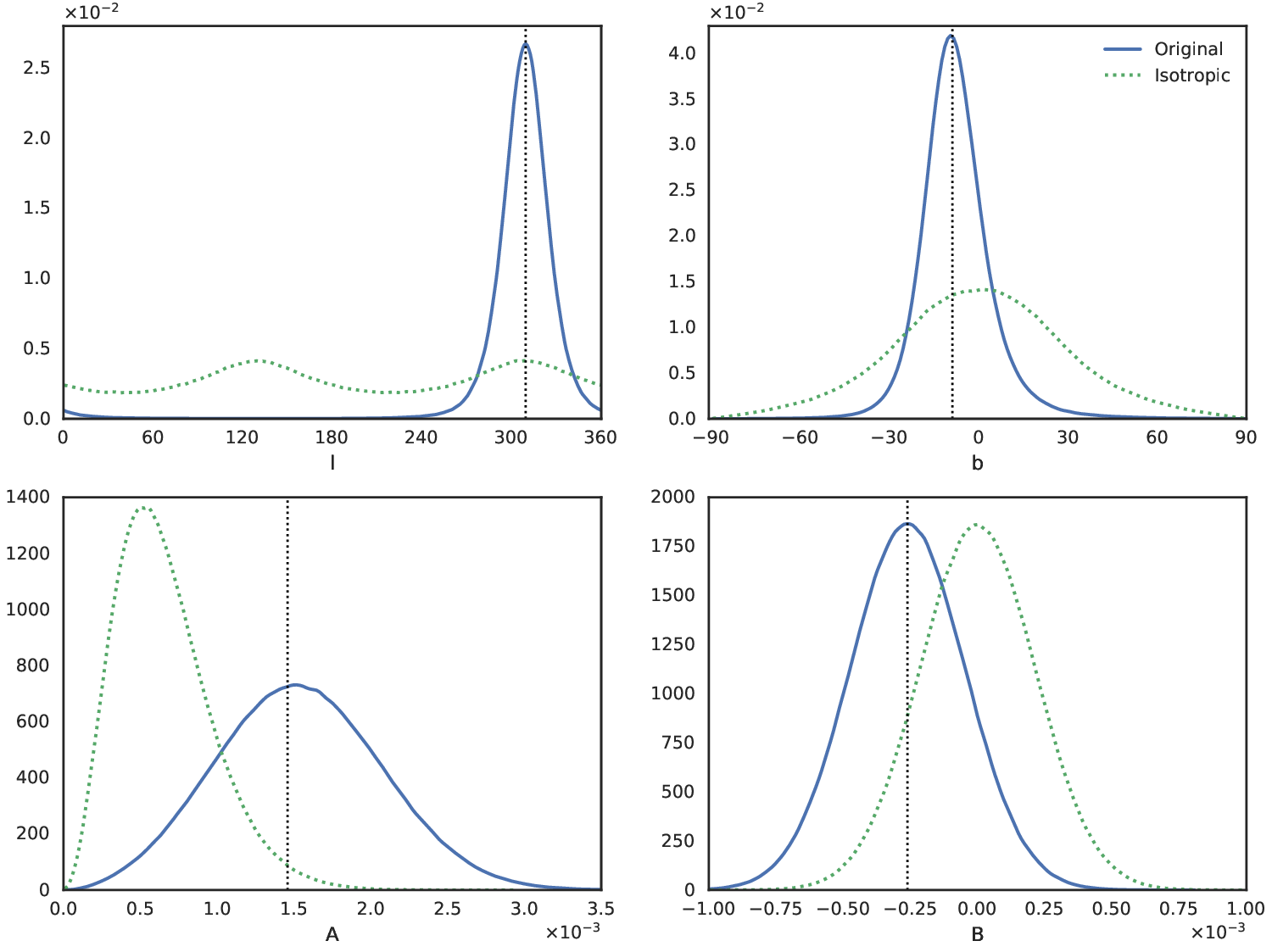}
   \caption{Blue lines show marginalized likelihoods of dipole $A$,
     monopole $B$ and $(l, b)$ for Union2.1 data set, and black vertical
     lines represent best-fitting values.  Green dotted lines represent
 results of ``isotropic'' samples.}\label{fig:Union_Hists}
\end{figure*}

\begin{figure*} \includegraphics[width=\textwidth]{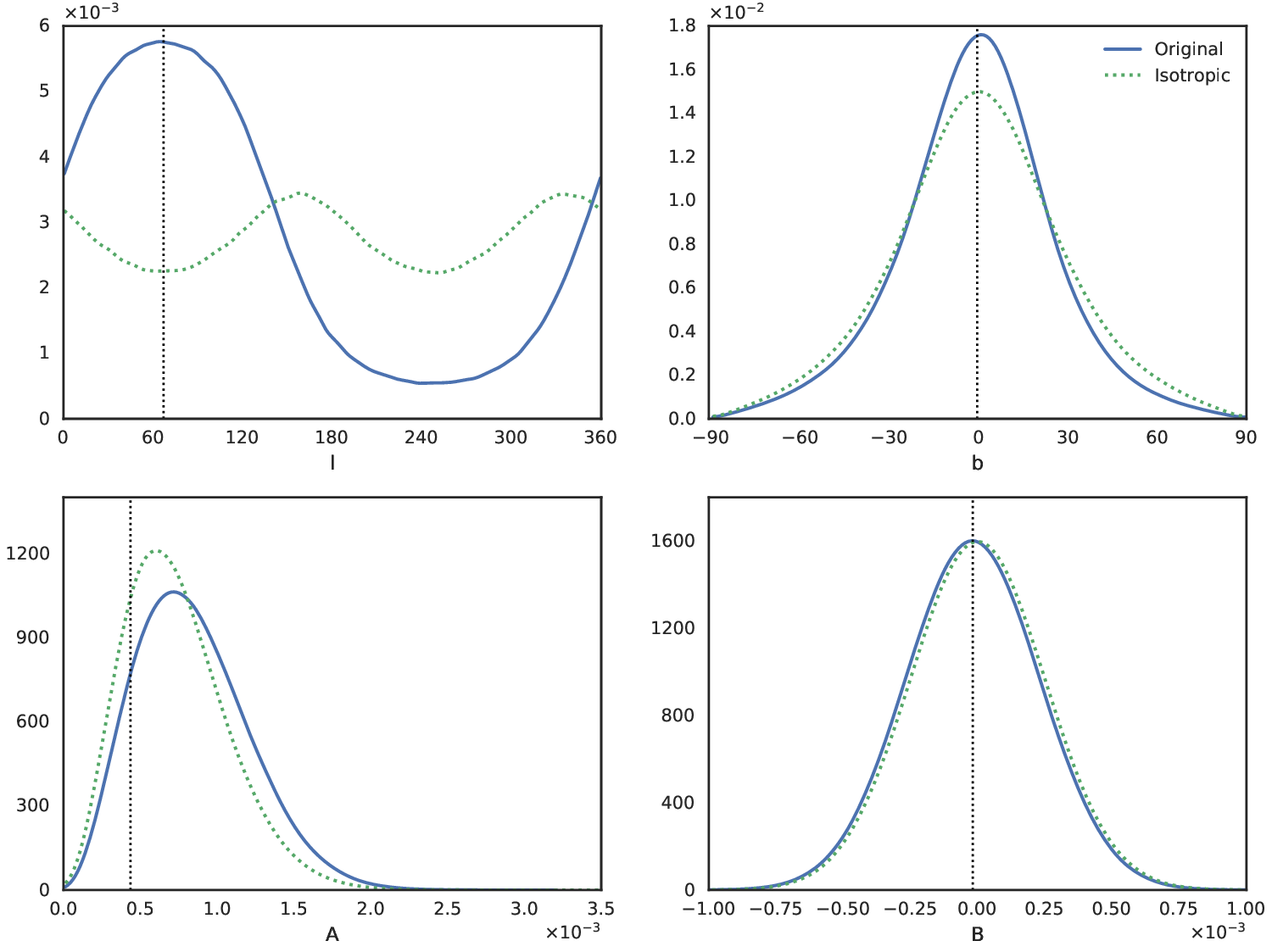}
 \caption{Similar to Figure \ref{fig:Union_Hists} for Constitution data
 set.  }\label{fig:Const_Hists} \end{figure*}

\begin{figure*}
   \includegraphics[width=\textwidth]{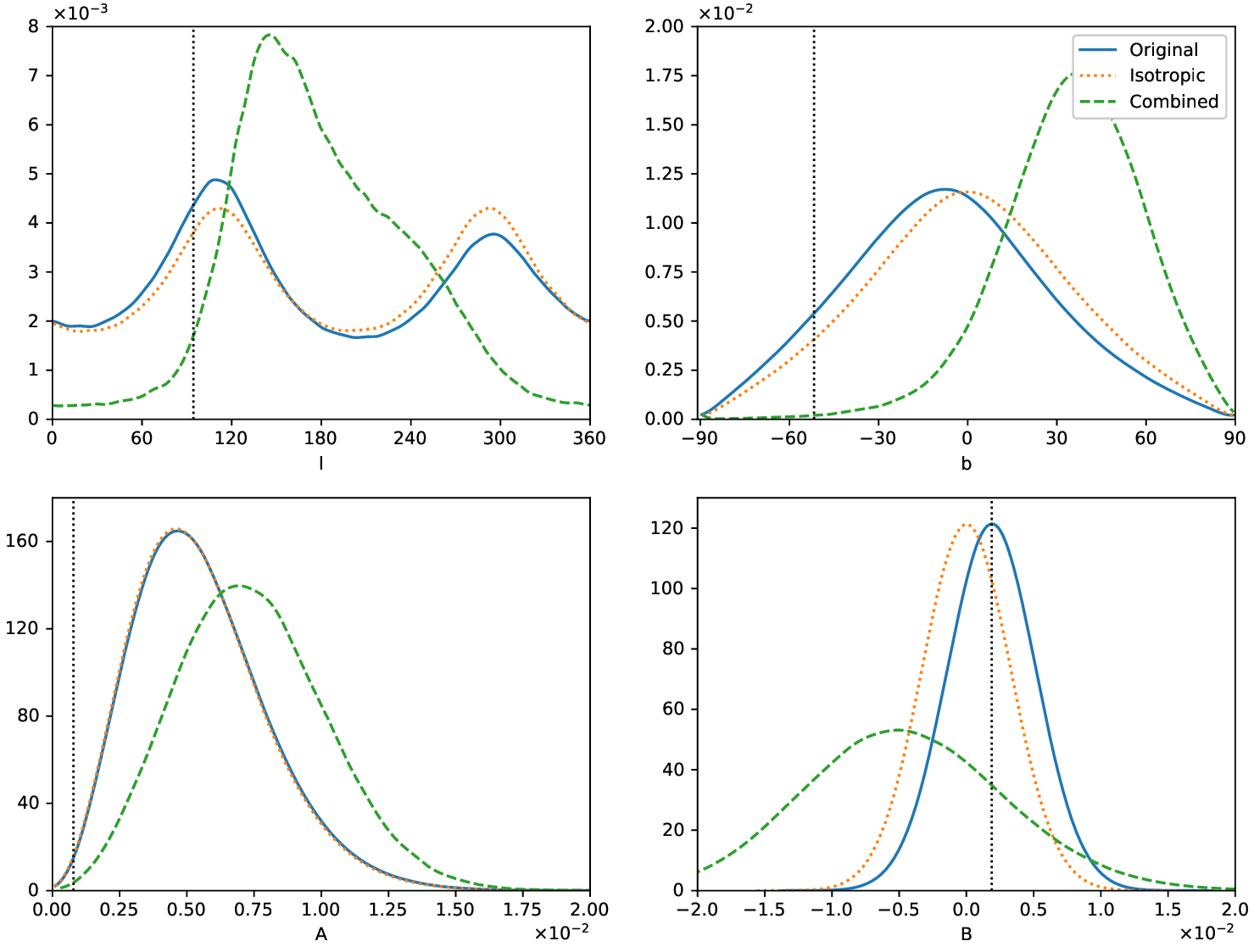}
   \caption{Blue lines show marginalized likelihoods of dipole $A$,
     monopole $B$ and $(l,b)$ for JLA data set, and black vertical lines
     represent best-fitting values.  Orange dotted lines represent
   results of ``isotropic'' samples.  Green dashed lines represent
 results of combined fitting of dipole parameters and nuisance
 parameters.  } \label{fig:JLA_combined_Hists} \end{figure*}

We generate $2 \times 10^6$ effective samples for each data set for
MCMC sampling. Probability distributions of dipole and monopole
parameters for Union2.1, Constitution, and JLA data sets are shown
in Figures \ref{fig:Union_Hists}, \ref{fig:Const_Hists} and
\ref{fig:JLA_combined_Hists}, respectively. Note that the
best-fitting parameters do not coincide with most probable values
for some parameters. This is because that the best-fitting values
represent the maximum point of probability density function (PDF),
while the most probable values in presented figures are the maximum
values of marginal PDF of $l, b, A$, respectively. The
marginalization process "flattens" lots of information about the
original PDF, thus the maxima can differ from those of the original
PDF. Also notice that because the volume of parameter space
approaches to zero when $A \to 0$ or $b \to \pm 90^{\circ}$, $A$
will always have zero likelihood at $A = 0$ while having its maximum
likelihood at some finite value, and $b$ will form a sine curve,
even when samples are uniformly distributed around the origin. This
does not affect the validity of analyzing the magnitude and
direction of the dipole, so long as the sampling results are
compared with those of ``isotropic'' samples, where deviations from
``isotropic'' scenario can be discerned from bias introduced from
features of the spherical coordinate.

For Union2.1 data set, the direction and magnitude of the dipole are
$(l=309.3^{\circ} {}^{+ 15.5^{\circ}}_{-15.7^{\circ}} ,\ b =
-8.9^{\circ} {}^{ + 11.2^{\circ}}_{-9.8^{\circ}})$, and  $\ A=(1.46 \pm
0.56) \times 10^{-3}$. The confidence level of dipolar anisotropy is
98.3\%. For Constitution data set, these parameters are
$(l=67.0^{\circ}{}^{+ 66.5^{\circ}}_{-66.2^{\circ}},\
b=-0.6^{\circ}{}^{+ 25.2^{\circ}}_{-26.3^{\circ}})$, $\ A=(4.4 \pm 5.0)
\times 10^{-4}$, and $\ B = (-0.2 \pm 2.4) \times 10^{-4}$. The
confidence level of dipolar anisotropy is 19.7\%.

\begin{figure*}
   \includegraphics[width=\textwidth]{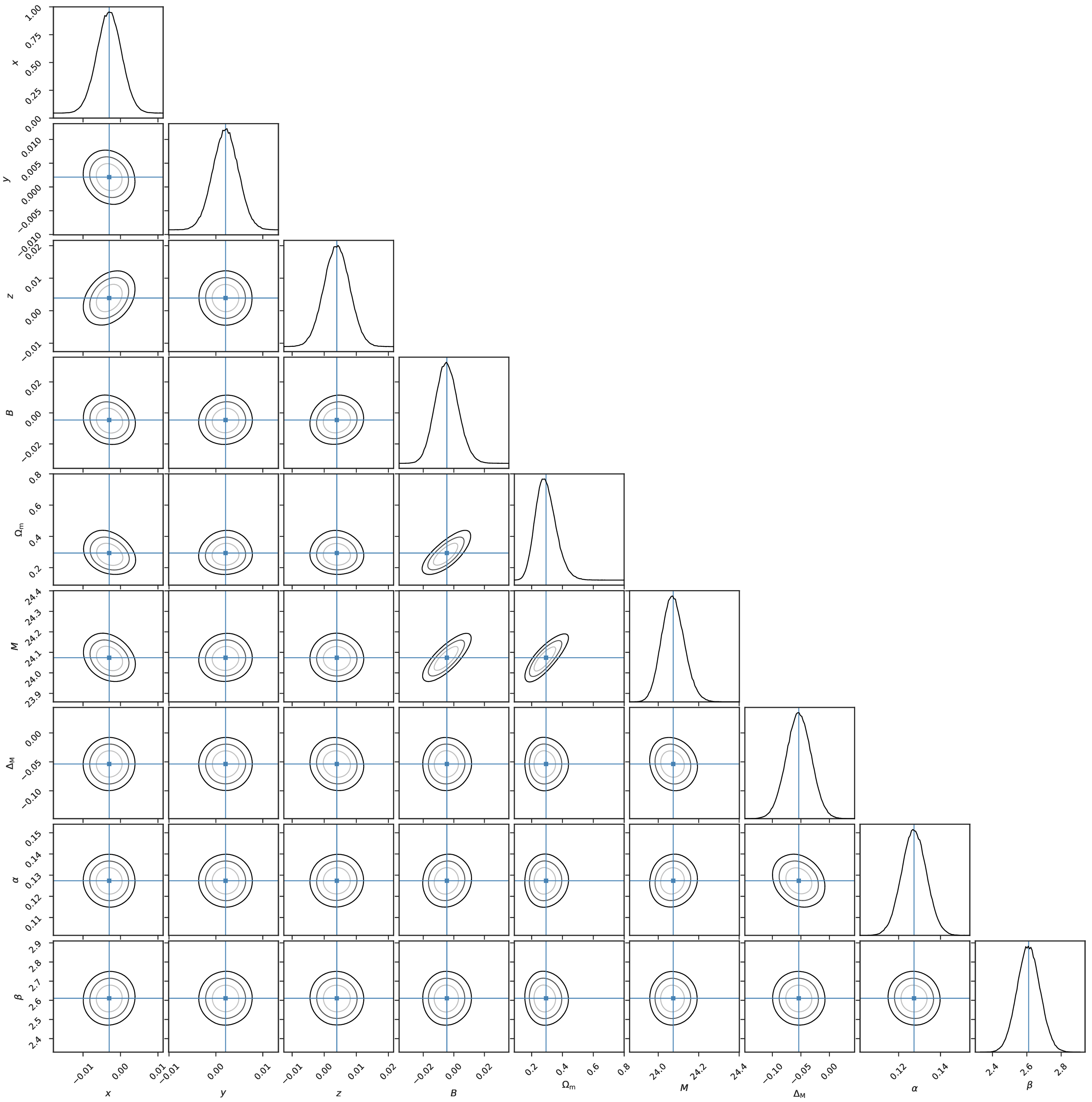}
   \caption{ Probability distribution functions of dipole parameters and
   nuisance parameters when fitted simultaneously for JLA data set.}
   \label{fig:JLA_combined} \end{figure*}

We fit JLA data set in two different approaches. Firstly, we fit
nuisance parameters, then fit dipole parameters using the fitted
nuisance parameters as constant. Secondly, we fit the nuisance
parameters and dipole parameters simultaneously. Since nuisance
parameters are related to both theoretical and observational values
of distance moduli, ``isotropic'' synthetic data sets are not
implemented in this approach. As is shown in Table
\ref{table:jla_method_compare}, the combined fitting approach gives
larger anisotropy than the separate fitting approach. Furthermore,
the fitted values of nuisance parameters are slightly shifted. As
shown in Figure \ref{fig:JLA_combined}, $B$, $\Omega_\mathrm{m}$, and
$M$ are correlated.

It is worth mentioning that, JLA data set gives null results in
dipole fitting. The 1$\sigma$ error range of dipole direction covers
the whole celestial sphere. The confidence level of dipole magnitude
is merely 0.23\%. Furthermore, there is no significant difference
between the likelihood of simulation results in 1$\sigma$ error range
and full results. The same is true for the likelihood of parameters
of ``isotropic'' samples and original samples. Besides, significant
deviations exist in best-fitting values and most probable values of
fitted parameters. Thus, no significant dipolar anisotropy of
redshift-distance modulus relation is found in the JLA data set.

For Union2.1 data set, we get similar results as previous works
\cite{Antoniou2010,Cai2012,Cai2013,Mariano2012,Yang2013,Wang2014,Lin2016}.
For Constitution data sets, our results are different from those of
\cite{Kalus2013}. Considering different methods, and the weak signal of
the dipole in this data set, the difference is reasonable.

For JLA data set, we get different likelihood distributions as
\cite{Lin2015}, which can be attributed to different fitting parameters
used in the MCMC estimation. In this paper, we fit the dipole by fitting
its rectangular components $(A_x, A_y, A_z)$, then convert the fitting
results to spherical coordinate. However, \cite{Lin2015} used the
galactic coordinate $(l, b)$ and dipole magnitude $A$ directly. The
likelihood distributions given in \cite{Lin2015} are inappropriate
because the marginalized likelihood of $b$ does not approach zero at $b
= \pm 90^{\circ}$.  Moreover, the marginalized likelihood of $l$ at
$0^{\circ}$ diverges from the likelihood at $360^{\circ}$, which is
contrary to the fact that the two longitudes actually ``wrap up'' on the
sphere.  We also notice that some other work, such as \cite{Chang2018,
Deng2018a, Deng2018b} shows similar improper likelihood distribution.

\begin{figure*} \includegraphics[width=\textwidth]{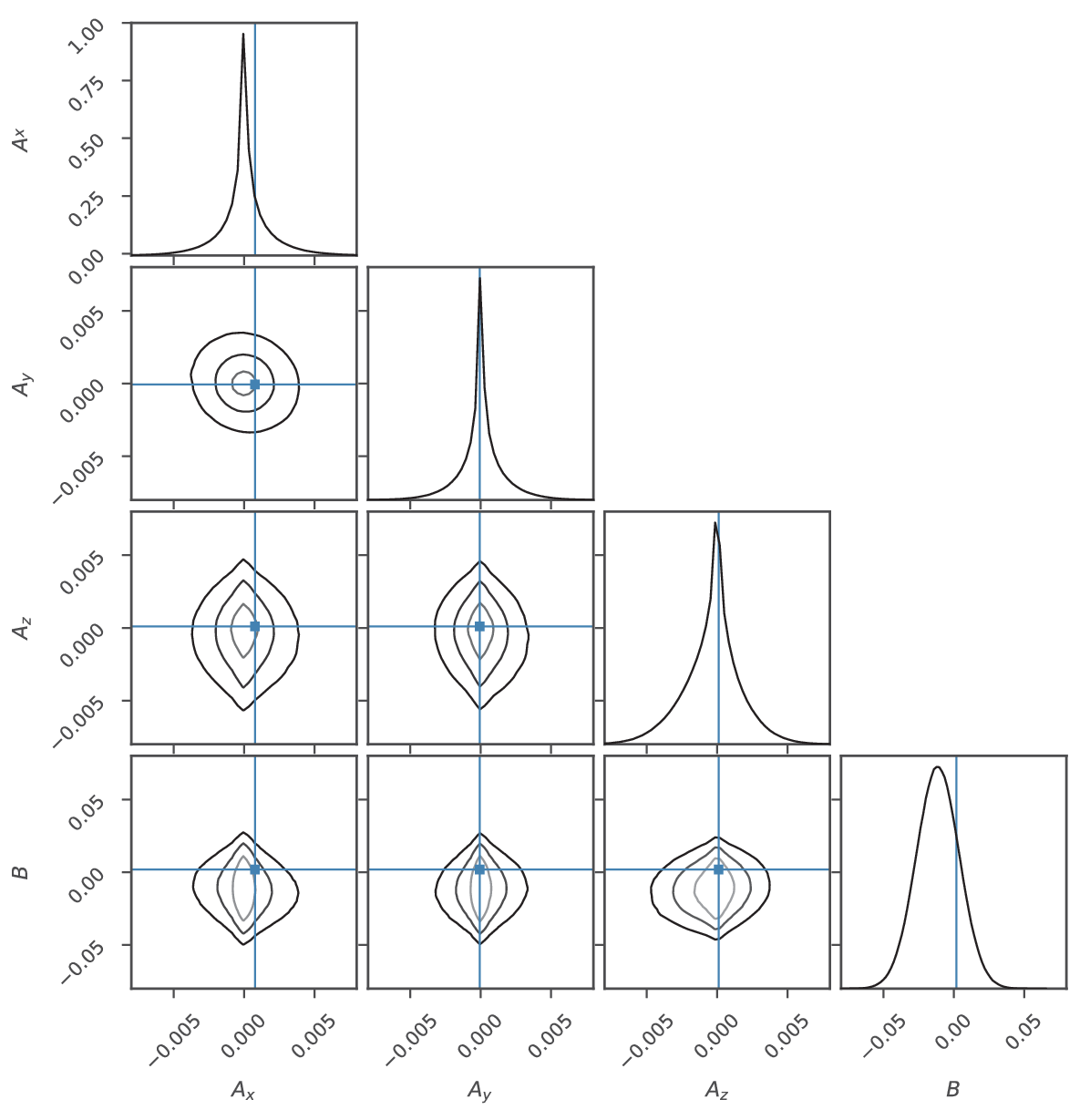}
   \caption{Probability distribution functions of rectangular components
     of dipole $\boldsymbol{A}$ and monopole $B$ for JLA data set
     directly using $(l,b,A)$ as fitted parameters.
 }\label{fig:JLA_A_xyz} \end{figure*}

\begin{figure*} \includegraphics[width=\textwidth]{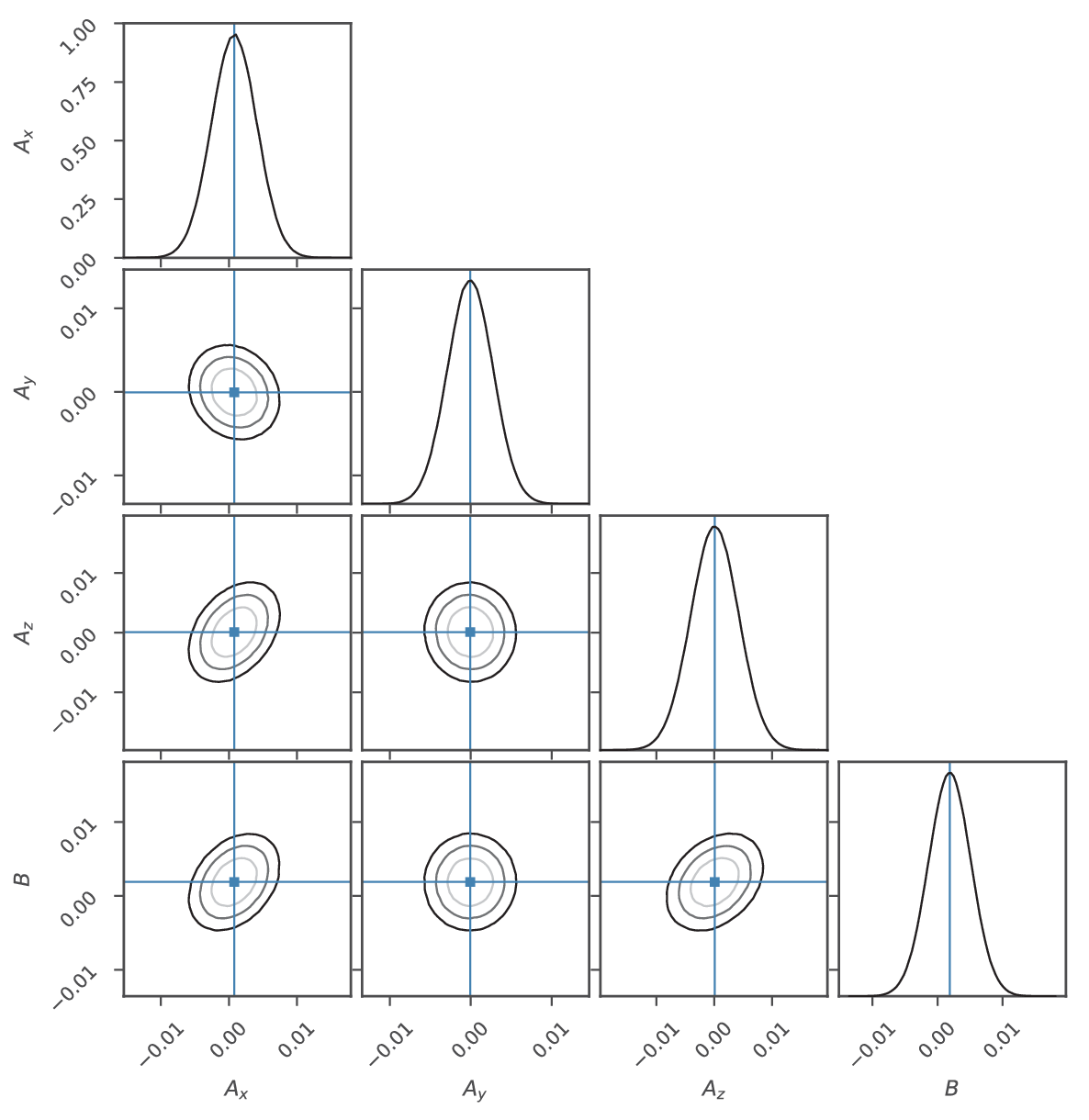}
   \caption{ Probability distribution functions of rectangular
   components of dipole $\boldsymbol{A}$ and monopole $B$ for JLA data
 set directly using $(A_x,A_y,A_z)$ as fitted
 parameters.}\label{fig:JLA_xyz} \end{figure*}

We fit parameter$(l, b, A)$ with constraints $0^{\circ} < l <
360^{\circ}, -90^{\circ} < b < 90^{\circ}, A > 0$ enforced on them,
i.e., sampling results out of such boundaries will be discarded by
the MCMC sampler. By using the fitting method described above, the
likelihood distributions mentioned in \cite{Lin2015} can be
reproduced. Though seemingly proper, such constraints ignore the
periodicity and symmetry of spherical coordinate, which results in
artificial ``gaps'' on fitting parameters, thus blocking the sampler
from properly sampling on those boundaries. As shown in Figure
\ref{fig:JLA_A_xyz}, non-zero likelihood at $b = \pm 90^{\circ}$
forms an unreasonable `spike' at poles, which depends on the choice
of the coordinate system. The proper method would be ``set free''
$(l, b, A)$ when sampling, then ``wrap around'' those parameters to
their boundaries afterward. By using this ``wrap around'' technique,
we reproduce the same posteriors as using rectangular components
$(A_x, A_y, A_z)$ for fitting. By comparison, posteriors used in this
paper are following best-fitting values, and joint likelihood
contours are smooth oval shapes, as shown in Figure
\ref{fig:JLA_xyz}.

\begin{figure*}
   \includegraphics[width=\textwidth]{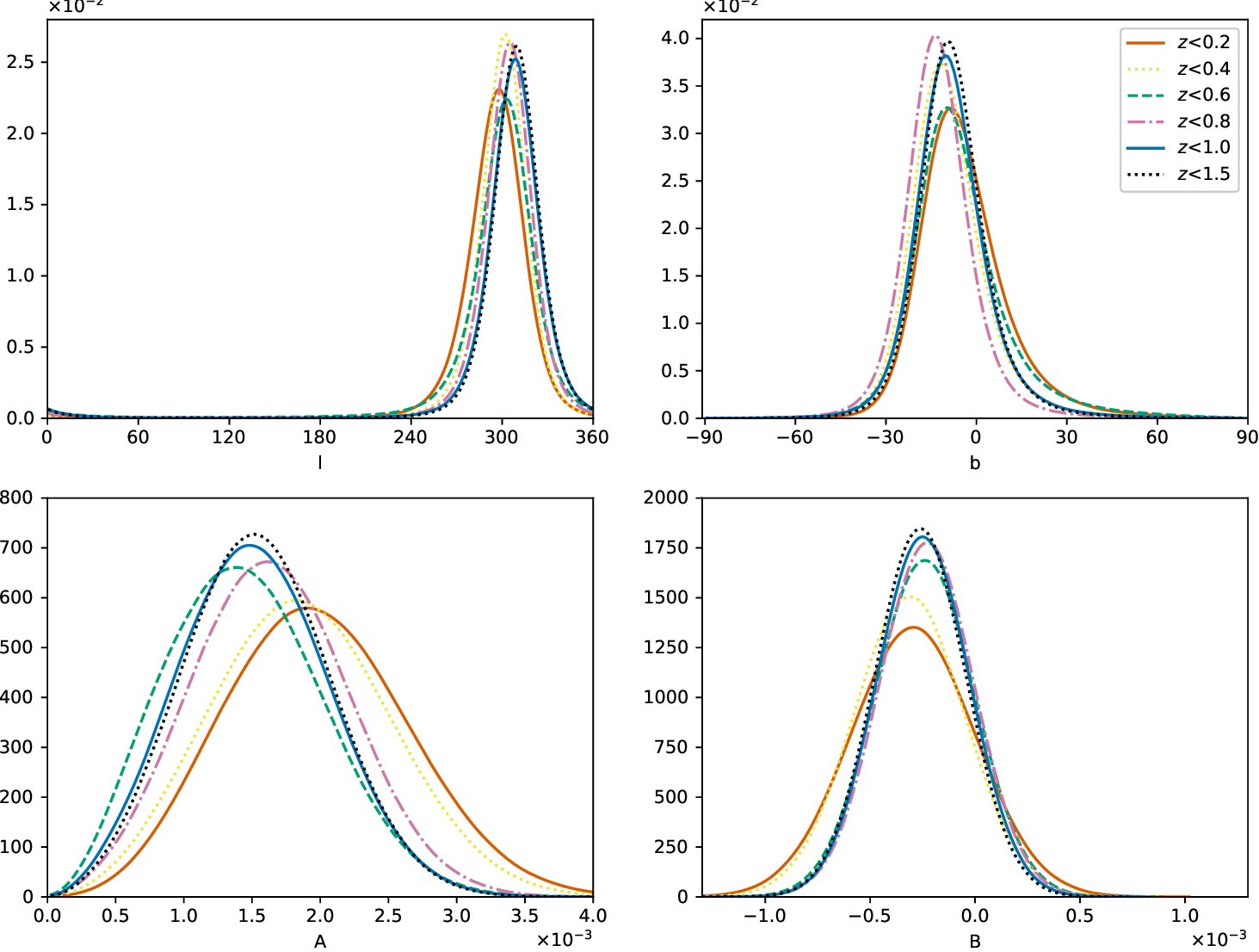}
   \caption{Marginalized likelihoods of dipole $A$, monopole $B$ and
   $(l,b)$ for different redshift range of Union2.1 data set.}
   \label{fig:UnionTomography} \end{figure*}

\begin{figure*}
 \includegraphics[width=\textwidth]{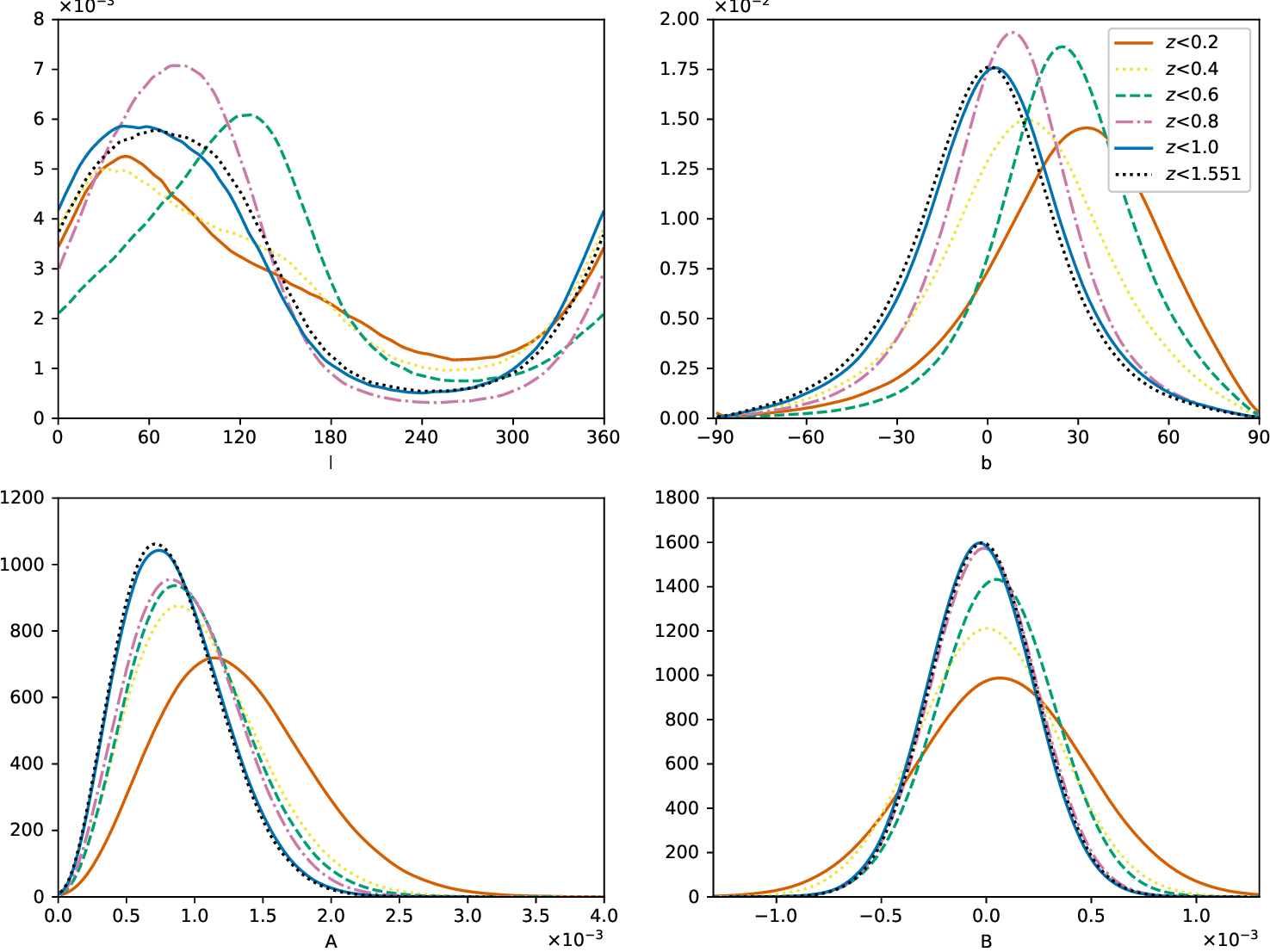}
 \caption{Similar to Figure \ref{fig:UnionTomography}, but for
 Constitution data set.} \label{fig:ConstTomography} \end{figure*}

The redshift tomography results for Union2.1 and Constitution data sets
are shown in Tables \ref{table:union_dipole_tomo} and
\ref{table:const_dipole_tomo}, respectively. The probability
distributions for each data set are shown in Figures
\ref{fig:UnionTomography} and \ref{fig:ConstTomography}. Because the
covariance matrix is involved in the fitting, it takes considerably long
time to give the results for JLA sample, we do not perform the redshift
tomography analysis.

\begin{figure*}
 \includegraphics[width=\textwidth]{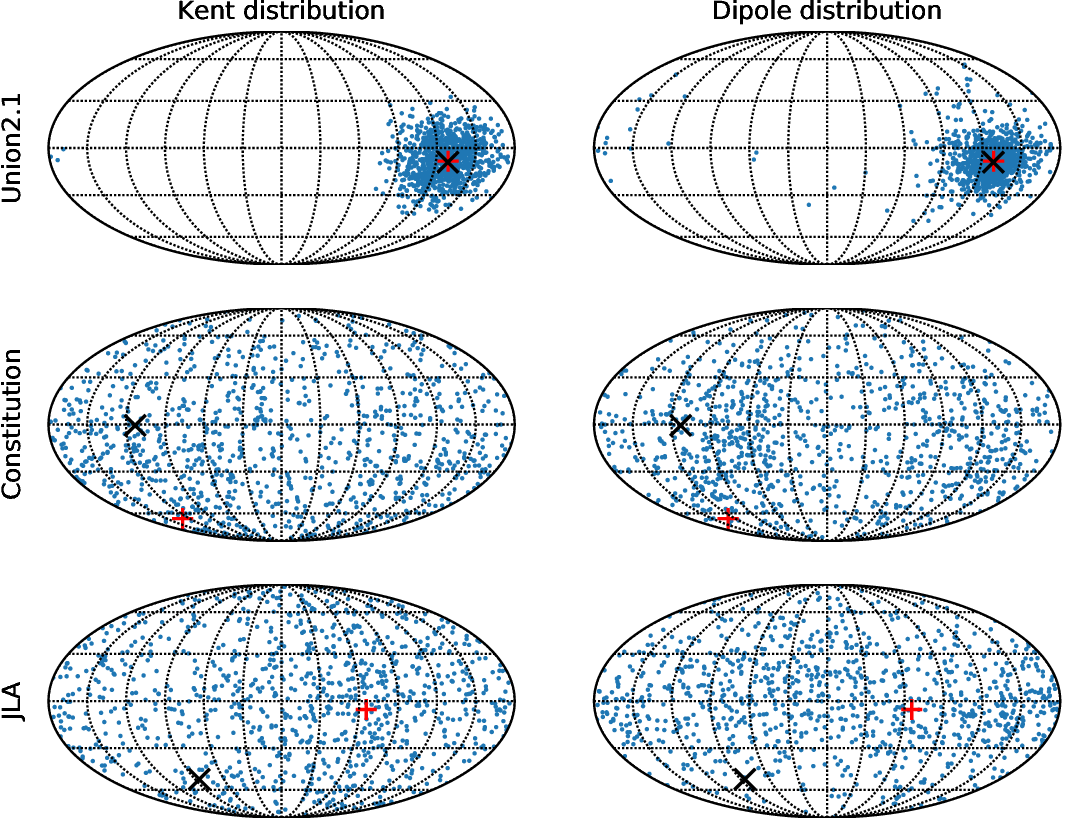}
 \caption{Left column shows samples drawn from fitted Kent
 distribution, the right column shows original dipole positions.
 Black x-cross shows the center of Kent distribution. Red cruciform
 shows the position of the best-fitting dipole.}
 \label{fig:Kent_dist_compare} \end{figure*}

We fit the spherical distribution of dipole position with Kent
distribution\cite{Kent1982}, which is the analogy to the bivariate normal
distribution on the two-dimensional unit sphere.  The samples drawn from
Kent distribution are shown along with the original dipole positions in
Figure \ref{fig:Kent_dist_compare}.  We find that Kent distribution fits
well with dipole positions derived from Union2.1 data set, but is less
suitable for Constitution and JLA data set.

\subsection{Quadrupole fitting method}

\begin{figure*}
 \includegraphics[width=\textwidth]{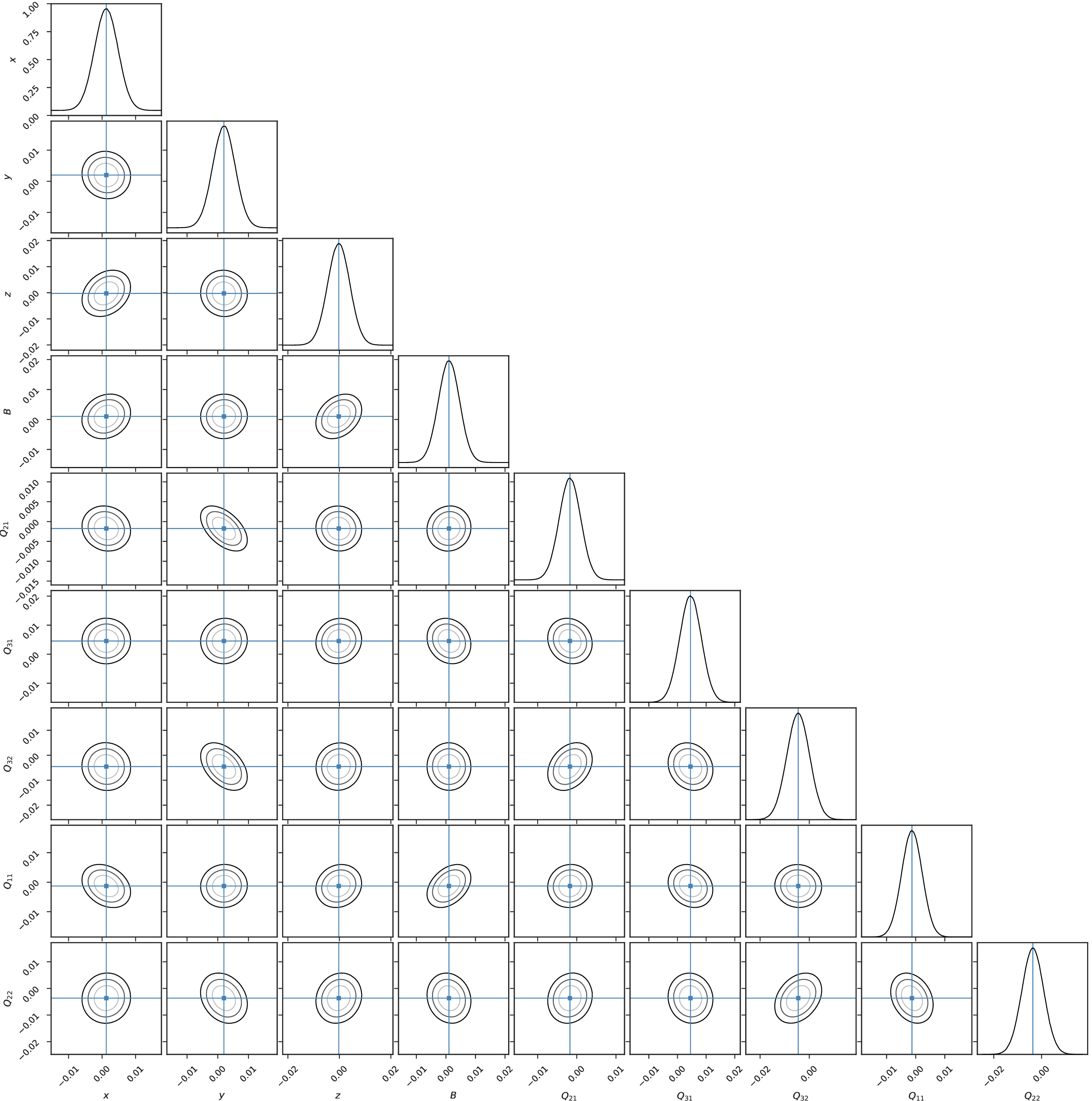} \caption{
 Probability distribution functions of dipole and quadrupole
 parameters.} \label{fig:JLA_quad_hist} \end{figure*}

\begin{figure*}
  \includegraphics[width=\textwidth]{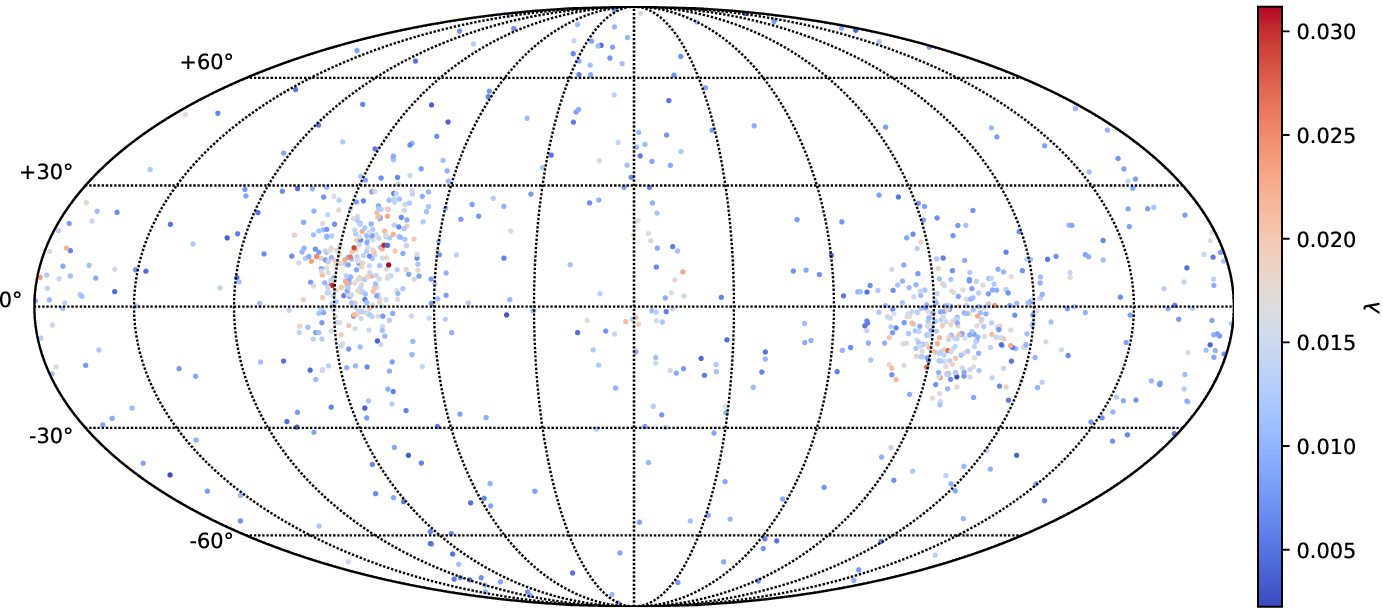}
  \caption{Distribution of eigenvectors with maximum eigenvalues of
  fitted quadrupole matrices, with eigenvalues represented in different
colors.} \label{fig:JLA_quad} \end{figure*}

We performed the quadrupole fitting method for the JLA data set. Five
independent parameters $Q_{11}, Q_{22}, Q_{12},\allowbreak Q_{13},
Q_{23}$ are used to represent the quadrupole $\mathbf{Q}$. The
likelihood distributions of these parameters are shown in Figure
\ref{fig:JLA_quad_hist}. The distributions of the main eigenvectors
are shown in Figure \ref{fig:JLA_quad}. The averaged absolute value
of the determinant of fitted matrices is $4.1 \times 10^{-7}$. The
quadrupole of JLA data set is not significant.

\subsection{Hemisphere comparison method}

The hemisphere comparison results in different redshift bins are
shown in Tables \ref{table:union_hemi_tomo} and
\ref{table:const_hemi_tomo}. The results are roughly following dipole
fitting results for Union2.1 data set but show larger discrepancies
for Constitution data set.

\section{Discussions}
\label{sec:discussions}

\subsection{Effects of anisotropy in data distribution}

As shown in Figures \ref{fig:Union_Hists}, \ref{fig:Const_Hists} and
\ref{fig:JLA_combined_Hists}, even if no redshift-distance anisotropy in
the input data, fitting results are still distributed an-isotropically
(green dotted lines).  This indicates other reasons, such as anisotropic
coordinate or redshift distribution would affect the results.


To determine whether the coordinate distribution of samples would
affect the results, we introduce two types of synthetic data sets.

\begin{description}
    \item[Type A data sets] substituting the coordinates in the original
      data set with random coordinates uniformly distributed in the
      whole sky. $\boldsymbol{\mu}_\mathrm{obs}$ are replaced with
      synthetic data.
    \item[Type B data sets] substituting the coordinates in the original
      data set with random coordinates, but only uniformly distributed
      in the eastern hemisphere of the celestial sphere. Distance moduli
      are substituted in the same manner as type A data sets.
\end{description}

\begin{figure*}
  \includegraphics[width=\textwidth]{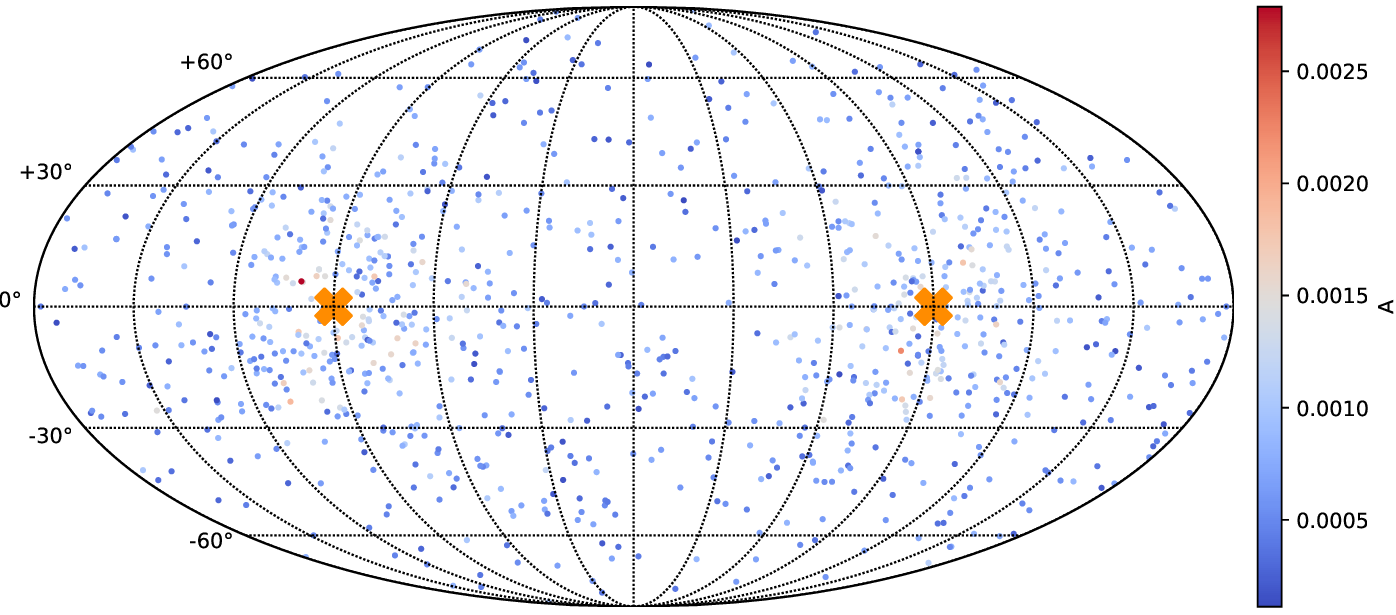}
  \caption{Distribution of dipole directions of type B samples. Dipole
  directions concentrate near crossed positions.}
  \label{fig:Union_sim_aniso_coords} \end{figure*}

\begin{figure*}
  \includegraphics[width=\textwidth]{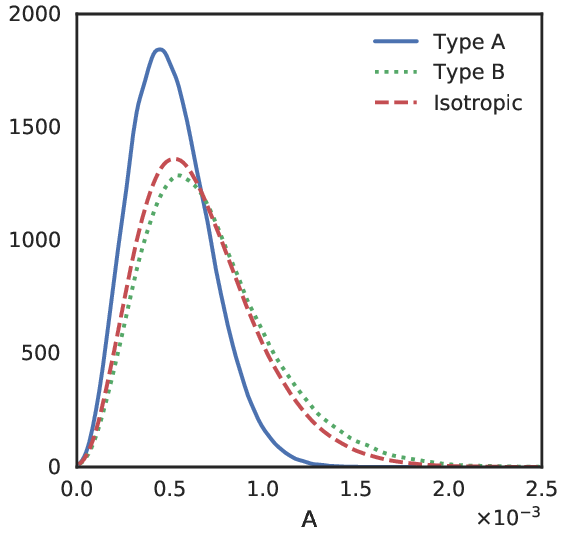}
  \caption{Likelihood of dipole magnitude $A$ of different samples.
  Blue solid line indicates the result of synthetic samples with
  isotropically distributed coordinates, which tend to reduce the
  dipole magnitude. Green dotted line indicates the result of synthetic
  samples with extremely an-isotropically distributed coordinates,
  which increase the dipole magnitude. }\label{fig:Union_simB}
\end{figure*}

Using MCMC sampling method introduced in section \ref{sec:methods}, we
find the dipoles in type A data sets are uniformly distributed in the
whole sky. However, the dipoles in Type B data sets tend to concentrate
in $(l,b)=(90^{\circ}, 0^{\circ})$, $(l,b)=(270^{\circ}, 0^{\circ})$, as
shown in Figure \ref{fig:Union_sim_aniso_coords}.  Meanwhile, dipole
magnitudes are generally larger than it is in Type A data sets, as shown
in Figure \ref{fig:Union_simB}. This indicates that the  anisotropy of
coordinates of samples does affect the results of dipole fitting.

\begin{figure*} \includegraphics[width=\textwidth]{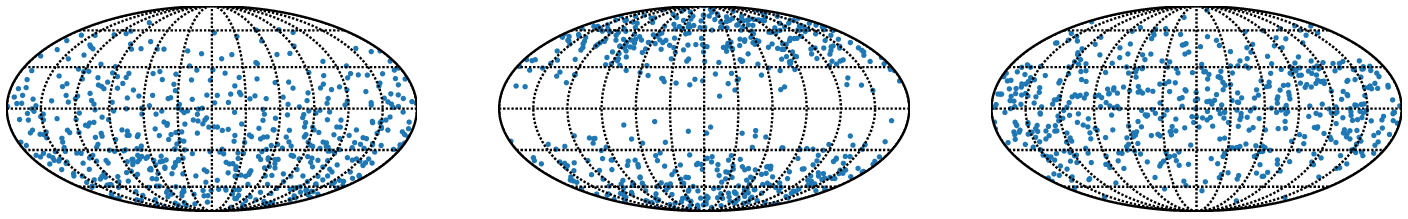}
  \caption{ Coordinates generated for three synthetic data sets.  The
    probability density functions of the coordinate density are
    proportional to $1-\sin(b)$, $\sin^2(b)$ and $\cos^2(b)$,
respectively.} \label{fig:aniso_coords} \end{figure*}

\begin{figure*}
   \includegraphics[width=\textwidth]{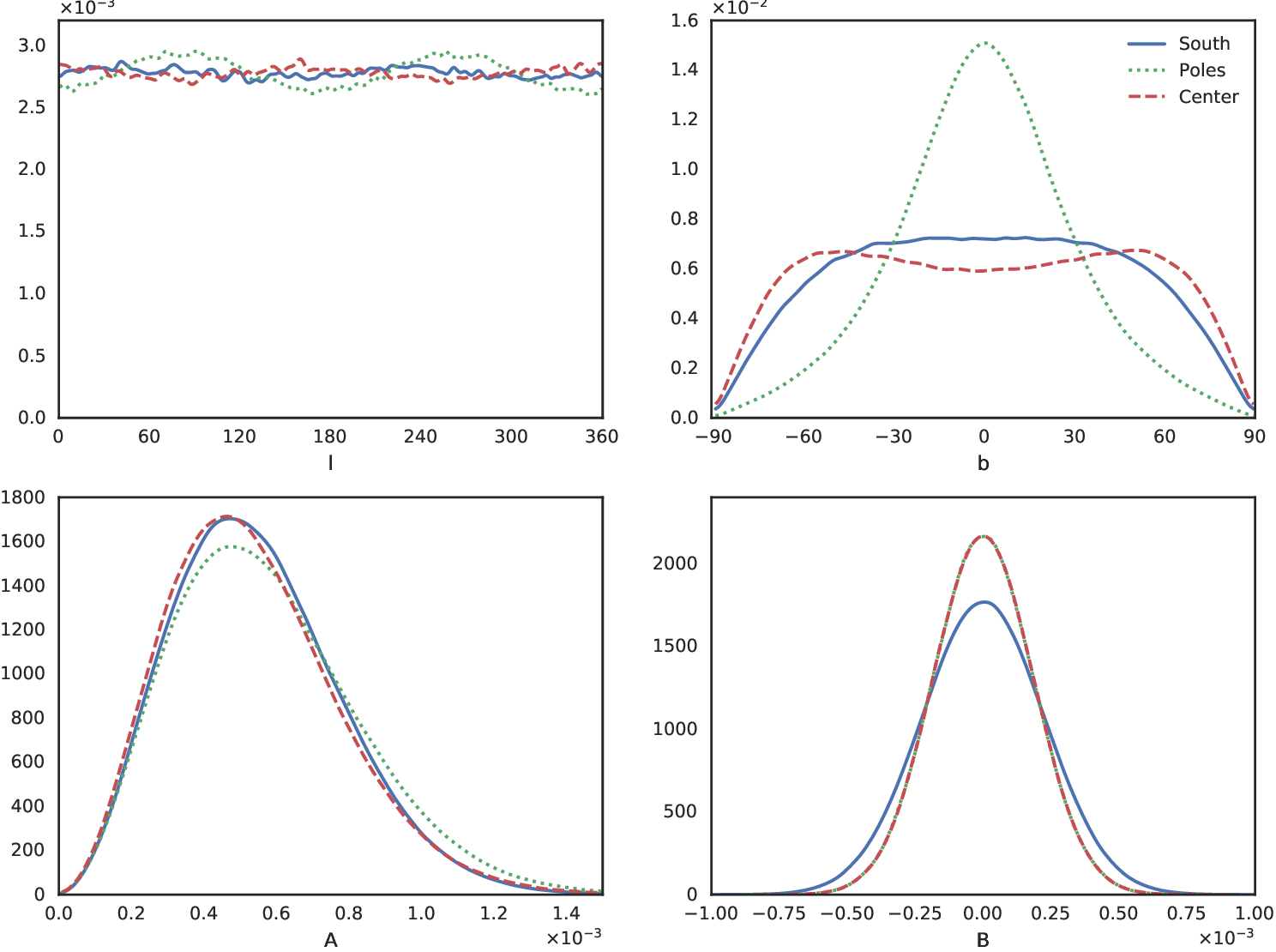}
   \caption{ Probability distributions of three synthetic data sets with
   a specific coordinate distribution.} \label{fig:Union_aniso_coords}
 \end{figure*}

In addition, we generate three kinds of synthetic data sets with
specific coordinate distributions, based on the Union2.1 data set.
The probability density function of the randomly-generated
coordinates are proportional to $1-\sin(b)$, $\sin^2(b)$ and
$\cos^2(b)$, respectively. In other words, the coordinates are
concentrated on the south galactic pole, both galactic poles, and the
galactic plane, respectively, as is shown in Figure
\ref{fig:aniso_coords}. Distance moduli are replaced in the same
manner as ``isotropic'' data sets. The fitted results are shown in
Figure \ref{fig:Union_aniso_coords}.




The spatial distribution of redshifts can be anisotropic, i.e., the
redshift of samples in one patch of the sky may be generally smaller
than another patch of sky. This may also cause an influence on fitting
results. To extract the effects of anisotropy in redshift distribution
from other factors, we introduce another kind of synthetic data set.

\begin{description}
 \item[Type C data sets] shuffle the coordinates in the original data
   set, making distance-related data of every sample correspond with a
   coordinate of another random sample in the data set.  Keep the
   `shuffled coordinates' order unchanged, then substitute distance
   moduli as type A data sets.
\end{description}

\begin{figure*}
  \includegraphics[width=\textwidth]{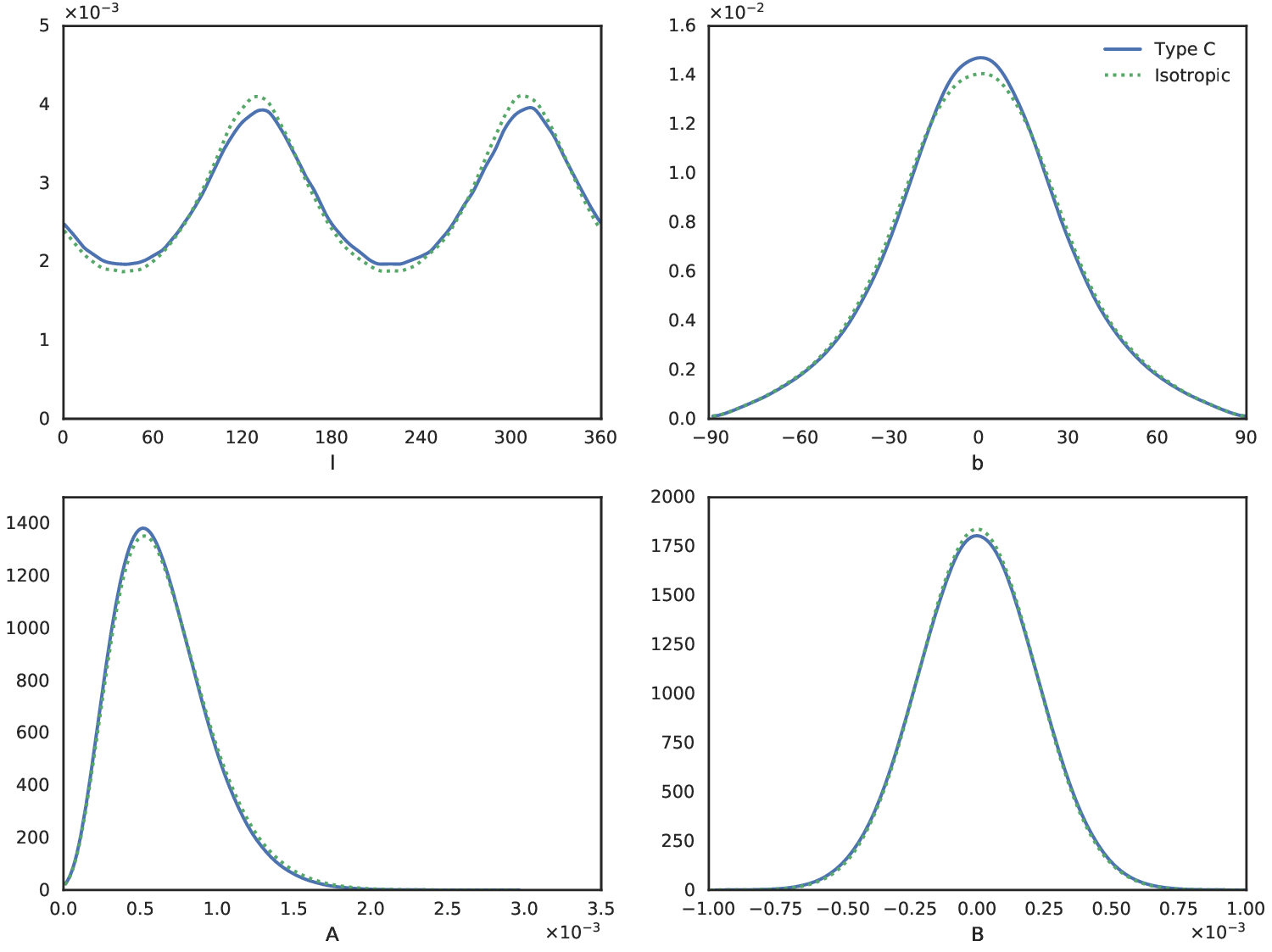}
  \caption{Likelihood of parameters for type C samples and isotropic
  samples. No significant deviation caused by the different spatial
distribution of redshifts is found.}\label{fig:Union_iso_hists}
\end{figure*}

Using the generating method described above, we can alternate the
spatial distribution of redshifts without changing the distribution of
coordinates. We find that the fitting results of type C data sets are
fairly consistent with ``isotropic'' samples, as shown in Figure
\ref{fig:Union_iso_hists}. Therefore, the anisotropy of redshift
distribution does not cause a significant influence on fitting results.

\section{Conclusions}
\label{sec:conclusions}

In this paper, we study three different data sets of SNe Ia, namely
Union2.1, Constitution, and JLA, to find possible dipolar anisotropy in
redshift-distance relation. We fit the dipole and monopole parameters by
minimizing $\chi^2$, then run MCMC sampling to determine the error range
and confidence level of the fitting parameters. We also apply
the hemisphere comparison method to find possible anisotropy in
$\Omega_\mathrm{m}$ on Union2.1 and Constitution data sets, and compare
the results with dipole fitting method.

For Union2.1 data set, we find the direction and magnitude of the dipole
are $(l=309.3^{\circ} {}^{+ 15.5^{\circ}}_{-15.7^{\circ}} ,\ b =
-8.9^{\circ} {}^{ + 11.2^{\circ}}_{-9.8^{\circ}}  ), \ A=(1.46 \pm 0.56)
\times 10^{-3}$. The monopole magnitude is $B = (-2.6 \pm 2.1) \times
10^{-4}$. The confidence level of dipolar anisotropy is 98.3\%.
Hemisphere comparison method gives
$\delta=0.20,l=352^{\circ},b=-9^{\circ}$. For Constitution data set,
these parameters are $(l=67.0^{\circ}{}^{+
66.5^{\circ}}_{-66.2^{\circ}},\ b=-0.6^{\circ}{}^{+
25.2^{\circ}}_{-26.3^{\circ}}), \ A=(4.4 \pm 5.0) \times 10^{-4},\ B =
(-0.2 \pm 2.4) \times 10^{-4}$,  The confidence level of dipolar
anisotropy is 19.7\%. The results of hemisphere comparison method
are$\delta=0.56,l=141^{\circ},b=-11^{\circ}$. For JLA data set, fitted
parameters are $l=94.4^{\circ},\ b=-51.7^{\circ}, \ A=7.8 \times
10^{-4},\ B=1.9 \times 10^{-3}$. The 1$\sigma$ error range of dipole
direction covers almost the whole sky. The confidence level of dipolar
anisotropy is merely 0.23\%.  Redshift tomography results show slightly
larger anisotropy at lower redshift range.

As shown above, the Union2.1 and Constitution data sets, although
have a large portion of overlapped data points, give radically
different results in terms of the best-fitting dipole parameters and
confidence levels. Furthermore, JLA data set, which contains more
data and a smaller portion of overlapped data than the other two
data sets, gives an essentially null result in dipolar anisotropy.
Due to the large discrepancies of the best-fitting dipole parameters
among different data sets, and the low confidence level given by
Constitution and JLA data set, we conclude that no sufficient
evidence of dipolar anisotropy is found in the aforementioned three
SNe Ia catalogs. The larger confidence level of dipolar anisotropy
shown in Union2.1 data set may come from non-cosmological factors.

We also study the effects of anisotropy of coordinate and redshift
distribution in dipole fitting method and find that anisotropy
distribution of coordinates can cause dipole magnitude to become
larger. However, anisotropy distribution of redshifts does not have a
significant influence on fitting results.

In future, the next-generation cosmological surveys, such as LSST
\cite{LSST2019}, Euclid \cite{Euclid2018}, and
WFIRST \cite{WFIRST2018} will observe much larger SNe Ia data
sets with enhanced light-curve calibration, which may shed light on the
anisotropy in redshift-distance relation.

\section*{Acknowledgments}

We thank the referee for helpful commentary on the manuscript. This work
is supported by the National Natural Science
Foundation of China (grant U1831207).


\begin{thebibliography}{}
\bibitem{Phillips1993}
M.M. Phillips, Astron. J. \textbf{413}, L105 (1993)

\bibitem{Riess1998}
A.G. Riess, et~al., Astron. J. \textbf{116}(3), 1009 (1998)

\bibitem{Perlmutter1999}
S.~Perlmutter, et~al., Astrophys. J. \textbf{517}(2), 565 (1999)

\bibitem{Kroupa2013}
P.~Kroupa, M.~Pawlowski, M.~Milgrom, Int. J. Mod. Phys. D \textbf{21}, 1230003
  (2012)

\bibitem{Kroupa2012}
P.~Kroupa, Publ. Astron. Soc. Aust. \textbf{29}(4), 395 (2012)

\bibitem{Perivolaropoulos2014}
L.~Perivolaropoulos, Galaxies \textbf{2}(1), 22 (2014)

\bibitem{Koyama2015}
K.~Koyama, Reports Prog. Phys. \textbf{79}, 46902 (2016)

\bibitem{Vielva2004}
P.~Vielva, et~al., Astrophys. J. \textbf{609}(1), 22 (2004)

\bibitem{DeOliveira-Costa2004}
A.~de~Oliveira-Costa, et~al., Phys. Rev. D \textbf{69}(6), 63516 (2004)

\bibitem{Tegmark2003}
M.~Tegmark, A.~{De Oliveira-Costa}, A.J.S. Hamilton, Phys. Rev. D
  \textbf{68}(12) (2003)

\bibitem{Hutsemekers2005}
D.~Hutsem{\'{e}}kers, et~al., Astron. Astrophys. \textbf{441}(3), 915 (2005)

\bibitem{Longo2009}
M.J. Longo, arXiv \textbf{04}, 1 (2009)

\bibitem{King2012}
J.A. King, et~al., Mon. Not. R. Astron. Soc. \textbf{422}(4), 3370 (2012)

\bibitem{Mariano2012}
A.~Mariano, L.~Perivolaropoulos, Phys. Rev. D \textbf{86}(8) (2012)

\bibitem{Campanelli2011}
{Campanelli, L}, et~al., Phys. Rev. D \textbf{83}(10) (2011)

\bibitem{Li2013}
X.~Li, et~al., Eur. Phys. J. C \textbf{73}, 2653 (2013)

\bibitem{Wang2018}
Y.Y. Wang, F.Y. Wang, Mon. Not. R. Astron. Soc. \textbf{474}(3), 3516 (2018)

\bibitem{Aluri2013}
P.K. Aluri, et~al., J. Cosmol. Astropart. Phys. \textbf{12}(1), 3 (2013)

\bibitem{Chang2014}
Z.~Chang, et~al., Eur. Phys. J. C \textbf{74}, 2821 (2014)

\bibitem{Antoniou2010}
I.~Antoniou, L.~Perivolaropoulos, J. Cosmol. Astropart. Phys. \textbf{12}, 12
  (2010)

\bibitem{Cai2012}
R.G. Cai, Z.L. Tuo, J. Cosmol. Astropart. Phys. \textbf{2012}, 6 (2012)

\bibitem{Cai2013}
R.G. Cai, et~al., Phys. Rev. D \textbf{87}(12), 123522 (2013)

\bibitem{Zhao2013}
W.~Zhao, P.~Wu, Y.~Zhang, Int. J. Mod. Phys. D \textbf{22}, 1350060 (2013)

\bibitem{Yang2013}
X.~Yang, F.Y. Wang, Z.~Chu, Mon. Not. R. Astron. Soc. \textbf{437}(2), 1840
  (2013)

\bibitem{Wang2014}
J.S. Wang, F.Y. Wang, Mon. Not. R. Astron. Soc. \textbf{443}(2), 1680 (2014)

\bibitem{Javanmardi2015}
B.~Javanmardi, et~al., Astrophys. J. \textbf{810}(1), 47 (2015)

\bibitem{BeltranJimenez2015}
J.B. Jimenez, V.~Salzano, R.~Lazkoz, Phys. Lett. B \textbf{741}, 168 (2015)

\bibitem{Lin2016}
H.N. Lin, X.~Li, Z.~Chang, Mon. Not. R. Astron. Soc. \textbf{460}(1), 617
  (2016)

\bibitem{Chang2018}
Z.~Chang, et~al., Mon. Not. R. Astron. Soc. \textbf{478}, 3633 (2018)

\bibitem{Deng2018a}
H.K. Deng, H.~Wei, Eur. Phys. J. C \textbf{78}, 755 (2018)

\bibitem{Deng2018b}
H.K. Deng, H.~Wei, Phys. Rev. D \textbf{97}, 123515 (2018)

\bibitem{Sun2018}
Z.Q. Sun, F.Y. Wang, Mon. Not. R. Astron. Soc. \textbf{478}, 5153 (2018)

\bibitem{Bonvin2006}
C.~Bonvin, R.~Durrer, M.~Kunz, Phys. Rev. Lett. \textbf{96}(19), 191302 (2006)

\bibitem{Schwarz2007}
D.J. Schwarz, B.~Weinhorst, Astron. Astrophys. \textbf{474}(3), 717 (2007)

\bibitem{Gordon2008}
C.~Gordon, K.~Land, A.~Slosar, Mon. Not. R. Astron. Soc. \textbf{387}(1), 371
  (2008)

\bibitem{Colin2011}
J.~Colin, et~al., Mon. Not. R. Astron. Soc. \textbf{414}(1), 264 (2011)

\bibitem{Turnbull2012}
S.~Turnbull, et~al., Mon. Not. R. Astron. Soc. \textbf{420}, 447 (2012)

\bibitem{Kalus2013}
B.~Kalus, et~al., Astron. Astrophys. \textbf{553}(120824), 56 (2013)

\bibitem{Appleby2014}
S.~Appleby, A.~Shafieloo, J. Cosmol. Astropart. Phys. \textbf{10}, 70 (2014)

\bibitem{Huterer2015}
D.~Huterer, D.~Shafer, F.~Schmidt, J. Cosmol. Astropart. Phys. \textbf{12}, 33
  (2015)

\bibitem{Cai2018}
R.G. Cai, et~al., Phys. Rev. D \textbf{97}, 103005 (2018)

\bibitem{Astier2006}
P.~Astier, et~al., Astron. Astrophys. \textbf{447}(1), 31 (2006)

\bibitem{Sullivan2011}
M.~Sullivan, et~al., Astrophys. J. \textbf{737}(2), 102 (2011)

\bibitem{Holtzman2008}
J.A. Holtzman, et~al., Astron. J. \textbf{136}(6), 2306 (2008)

\bibitem{Rest2013}
A.~Rest, et~al., Astrophys. J. \textbf{795}(1), 44 (2014)

\bibitem{Scolnic2013}
D.M. Scolnic, et~al., Astrophys. J. \textbf{780}(1), 37 (2014)

\bibitem{Tonry2012}
J.L. Tonry, et~al., Astrophys. J. \textbf{750}(2), 99 (2012)

\bibitem{Hicken2009}
M.~Hicken, et~al., Astrophys. J. \textbf{700}(2), 1097 (2009)

\bibitem{Stritzinger2011}
M.~Stritzinger, et~al., Astron. J. \textbf{142}(5) (2011)

\bibitem{Folatelli2009}
G.~Folatelli, et~al., Astron. J. \textbf{139}(1), 120 (2009)

\bibitem{Contreras2009}
C.~Contreras, et~al., Astron. J. \textbf{139}(2), 519 (2009)

\bibitem{Ganeshalingam2013}
M.~Ganeshalingam, W.~Li, A.V. Filippenko, Mon. Not. R. Astron. Soc.
  \textbf{433}(3), 2240 (2013)

\bibitem{Aldering2002}
G.~Aldering, et~al., in \emph{Proc. SPIE}, vol. 4836, ed. by J.A. Tyson,
  S.~Wolff (2002), vol. 4836, pp. 61--72

\bibitem{Rubin2009}
D.~Rubin, et~al., Astrophys. J. \textbf{695}(1), 391 (2009)

\bibitem{Campbell2013}
H.~Campbell, et~al., Astrophys. J. \textbf{763}(2), 88 (2013)

\bibitem{Conley2011}
A.~Conley, et~al., Astrophys. Journal, Suppl. \textbf{192}(1), 1 (2011)

\bibitem{Suzuki2012}
N.~Suzuki, et~al., Astrophys. J. \textbf{746}(1), 85 (2012)

\bibitem{Betoule2014}
M.~Betoule, et~al., Astron. Astrophys. \textbf{568}, A22 (2014)

\bibitem{Guillochon2016}
{Guillochon, James}, et~al., Astrophys. J. \textbf{835}(1) (2016)

\bibitem{Visser2004}
M.~Visser, Class. Quantum Gravity \textbf{21}(11), 2603 (2004)

\bibitem{Cattoen2007}
C.~Catto{\"{e}}n, M.~Visser, Class. Quantum Gravity \textbf{24}, 5985 (2007)

\bibitem{Wang2009}
F.Y. Wang, Z.G. Dai, S.~Qi, Astron. Astrophys. \textbf{507}(1), 53 (2009)

\bibitem{Johansson2013}
J.~Johansson, et~al., Mon. Not. R. Astron. Soc. \textbf{435}(2), 1680 (2013)

\bibitem{Lin2015}
H.N. Lin, et~al., Mon. Not. R. Astron. Soc. \textbf{456}(2), 1881 (2015)

\bibitem{Kent1982}
J.T. Kent, J. R. Stat. Soc. Ser. B \textbf{44}(1), 71 (1982)

\bibitem{LSST2019}
{\v{Z}}.~Ivezi{\'{c}}, et~al., Astrophys. J. \textbf{873}, 111 (2019)

\bibitem{Euclid2018}
L.~Amendola, et~al., Living Rev. Relativ. \textbf{21}, 2 (2018)

\bibitem{WFIRST2018}
R.~Hounsell, et~al., Astrophys. J. \textbf{867}, 23 (2018)

\end{thebibliography}

\newpage

\begin{table*}
\tiny \caption{Incomplete list of previous works on cosmological
preferred directions. AM is short for a specific anisotropic model,
DF for dipole fitting method, and HC for hemisphere comparison
method.}
\label{table:previous-works}
\begin{tabular}{llllll}
    \hline
Authors                                                & Sample Used     & Method & $(l,b)$                                                                                        & Anisotropy Level                             & C.L.         \\
    \hline
Antoniou \& Perivolaropoulos (2010)\cite{Antoniou2010} & Union2          & HC     & $(306^{\circ}, 15^{\circ})$                                                                    & $\Delta \Omega_{m} / \Omega_{m} = 0.42$      & 70\%         \\
Cai \& Tuo (2012)\cite{Cai2012}                        & Union2          & HC     & $(314^{\circ} {}^{+ 20^{\circ}}_{-13^{\circ}} , 28^{\circ} {}^{ +11^{\circ}}_{-33^{\circ}}  )$ & $\Delta q_0/q_0 = 0.79 {}^{+0.27}_{-0.28}$   &              \\
Cai et al. (2013)\cite{Cai2013}                        & Union2+67GRB    & AM     & $(306^{\circ}, -13^{\circ})$                                                                   & $g_0 = 0.030 {}^{+0.010}_{-0.030}$           & 2$\sigma$    \\
Mariano \& Perivolaropoulos (2012)\cite{Mariano2012}   & Union2          & DF     & $(309.4^{\circ} \pm 18.0^{\circ} , -15.1^{\circ} \pm 11.5^{\circ} )$                           & $A=(1.3 \pm 0.6) \times 10^{-3}$             & 2$\sigma$    \\
                                                       & Keck+VLT        & DF     & $(320.5^{\circ} \pm 11.8^{\circ} , -11.7^{\circ} \pm 7.5^{\circ} )$                            & $A=(1.02 \pm 0.25) \times 10^{-5}$           & 3.9 $\sigma$ \\
Kalus et al. (2013)\cite{Kalus2013}                    & Constitution    & HC     & $(-35^{\circ}, -19^{\circ})$                                                                   & $\Delta H/H \sim 0.026$                      & 95\%         \\
Yang et al. (2013)\cite{Yang2013}                      & Union2.1        & HC     & No Significance                                                                                & No Significance                              &              \\
                                                       &                 & DF     & $(307.1^{\circ} \pm 16.2^{\circ} , -14.3^{\circ} \pm 10.1^{\circ} )$                           & $A=(1.2 \pm 0.5) \times 10^{-3}$             & 95.45\%      \\
Wang \& Wang (2014)\cite{Wang2014}                     & Union2.1+116GRB & DF     & $(309.2^{\circ} \pm 15.8^{\circ} , -8.6^{\circ} \pm 10.5^{\circ} )$                            & $A=(1.37 \pm 0.57) \times 10^{-3}$           & 97.29\%      \\
Lin et al. (2015)\cite{Lin2015}                        & JLA             & DF     & No Significance                                                                                & No Significance                              &              \\
Lin et al. (2016)\cite{Lin2016}                        & Union2.1        & HC     & $(241.9^{\circ}, -19.5^{\circ})$                                                               & $\Delta \Omega_{m} / \Omega_{m} = 0.306$     & 37\%         \\
                                                       &                 & AM     & $(310,6^{\circ} \pm 18.2^{\circ} , -13.0^{\circ} \pm 11.1^{\circ} )$                           & $D = (1.2 \pm 0.5) \times 10^{-3}$           & 95.9\%       \\
Deng \& Wei (2018a)\cite{Deng2018a}                    & JLA             & DP     & $(309.4^{\circ} \pm 18.0^{\circ} , -15.1^{\circ} \pm 11.5^{\circ} )$                           & $A=(1.3 \pm 0.6) \times 10^{-3}$             & 2$\sigma$    \\
                                                       &                 & HC     & $(23^{\circ}, 2^{\circ}) \& (299^{\circ}, 28^{\circ})$                                         & $\Delta \Omega_{m} / \Omega_{m} = 0.31,0.28$ &              \\
Deng \& Wei (2018b)\cite{Deng2018b}                    & Pantheon        & DF     & No Significance                                                                                & No Significance                              &              \\
                                                       &                 & HC     & $(138^{\circ}, -7^{\circ}) \& (102^{\circ}, -29^{\circ})$                                      & $\Delta \Omega_{m} / \Omega_{m} = 0.30,0.24$ &              \\
Sun \& Wang (2018)\cite{Sun2018}                       & Pantheon        & DF     & $(108^{\circ}{}^{+ 43^{\circ}}_{-77^{\circ}}, 7^{\circ}{}^{+ 41^{\circ}}_{-77^{\circ}})$       & $A=(2.6 \pm 2.6) \times 10^{-4}$             & 36.2\%       \\
                                                       &                 & HC     & $(110^{\circ} \pm 11^{\circ} , 15^{\circ}\pm 19^{\circ} )$                                     & $\Delta \Omega_{m} / \Omega_{m} = 0.105$     &              \\
    \hline
\end{tabular}
\end{table*}

\begin{table*}
 \centering
\caption{Best-fitting results of dipole and monopole for three data sets. The
confidence level (C.L.) is defined as the probability of the case where the
dipole magnitude of an arbitrary data set in ``isotropic'' samples being
smaller than the best-fitting dipole magnitude. }
    \label{table_results}
\begin{tabular}{l|rrr}
    \hline
data sets & Union 2.1                                           & Constitution                                      & JLA                  \\
    \hline
$l$      & $309.3^{\circ} {}^{+ 15.5^{\circ}}_{-15.7^{\circ}}$ & $67.0^{\circ}{}^{+ 66.5^{\circ}}_{-66.2^{\circ}}$ & $94.4^{\circ}$       \\
$b$      & $-8.9^{\circ} {}^{ + 11.2^{\circ}}_{-9.8^{\circ}}$  & $-0.6^{\circ}{}^{+ 25.2^{\circ}}_{-26.3^{\circ}}$ & $-51.7^{\circ}$      \\
$A$      & $(1.46 \pm 0.56) \times 10^{-3}$                    & $(4.4 \pm 5.0) \times 10^{-4}$                    & $7.8 \times 10^{-4}$ \\
$B$      & $(-2.6 \pm 2.1) \times 10^{-4}$                     & $(-0.2 \pm 2.4) \times 10^{-4}$                   & $1.9 \times 10^{-3}$ \\
C.L.     & 98.3\%                                              & 19.7\%                                            & 0.23\%               \\
    \hline
\end{tabular}
\end{table*}

\begin{table*}
 \centering
    \caption{Fitted parameters of JLA data set for separated fitting and combined fitting.}
    \label{table:jla_method_compare}
\begin{tabular}{l|rrrr|rrrrr}
      \hline
Fitting method &           $l$ &           $b$ &                  $A$ &                   $B$ & $M_\mathrm{B}^1$ & $\Delta_M$ & $\alpha$ & $\beta$ & $\Omega_\mathrm{m}$ \\
      \hline
   Separated & $ 94.4^\circ$ & $-51.7^\circ$ & $7.8 \times 10^{-4}$ &  $1.9 \times 10^{-3}$ &            24.12 &      -0.06 &    0.140 &    3.12 &                0.29 \\
    Combined & $171.7^\circ$ & $ 48.2^\circ$ & $5.3 \times 10^{-3}$ & $-4.6 \times 10^{-3}$ &            24.07 &      -0.05 &    0.127 &    2.61 &                0.29 \\
      \hline
\end{tabular}
\end{table*}

\begin{table*}
 \centering
    \caption{Redshift tomography results for the Union2.1 data set using dipole fitting method.}
    \label{table:union_dipole_tomo}
\begin{tabular}{l|rrrr}
      \hline
$z$ range    &             $l$ &              $b$ &                  $A$ &                  $B$ \\
      \hline
     $z<0.2$ & $296.9^{\circ}$ & $-6.3  ^{\circ}$ & $1.8 \times 10^{-3}$ & $-3.0\times 10^{-4}$ \\
     $z<0.4$ & $301.2^{\circ}$ & $-11.0 ^{\circ}$ & $1.8 \times 10^{-3}$ & $-3.1\times 10^{-4}$ \\
     $z<0.6$ & $302.2^{\circ}$ & $-8.1  ^{\circ}$ & $1.3 \times 10^{-3}$ & $-2.4\times 10^{-4}$ \\
     $z<0.8$ & $304.8^{\circ}$ & $-13.5 ^{\circ}$ & $1.6 \times 10^{-3}$ & $-2.3\times 10^{-4}$ \\
     $z<1.0$ & $308.5^{\circ}$ & $-9.7  ^{\circ}$ & $1.4 \times 10^{-3}$ & $-2.5\times 10^{-4}$ \\
     $z<1.5$ & $309.4^{\circ}$ & $-8.9  ^{\circ}$ & $1.5 \times 10^{-3}$ & $-2.6\times 10^{-4}$ \\
      \hline
\end{tabular}
\end{table*}

\begin{table*}
 \centering
 \caption{Similar to Table \ref{table:union_dipole_tomo}, but for the Constitution data set.}
    \label{table:const_dipole_tomo}
\begin{tabular}{l|rrrr}
        \hline
$z$ range      &             $l$ &            $b$ &                  $A$ &                    $B$ \\
        \hline
       $z<0.2$ &  $64.1^{\circ}$ & $54.5^{\circ}$ & $6.9 \times 10^{-4}$ &  $7.1 \times 10^{-5} $ \\
       $z<0.4$ &  $65.4^{\circ}$ & $28.3^{\circ}$ & $4.2 \times 10^{-4}$ &  $6.9 \times 10^{-6} $ \\
       $z<0.6$ & $106.0^{\circ}$ & $41.9^{\circ}$ & $5.9 \times 10^{-4}$ &  $4.6 \times 10^{-5} $ \\
       $z<0.8$ &  $73.3^{\circ}$ & $10.6^{\circ}$ & $5.9 \times 10^{-4}$ & $-1.2 \times 10^{-5} $ \\
       $z<1.0$ &  $59.9^{\circ}$ & $3.1 ^{\circ}$ & $4.5 \times 10^{-4}$ & $-2.8 \times 10^{-5} $ \\
     $z<1.55$ &  $67.0^{\circ}$ & $-0.6^{\circ}$ & $4.4 \times 10^{-4}$ &  $-1.8 \times 10^{-5}$ \\
        \hline
\end{tabular}
\end{table*}

\begin{table*}
 \centering
    \caption{Redshift tomography results using hemisphere comparison method for the Union2.1 data set.}
    \label{table:union_hemi_tomo}
\begin{tabular}{l|rrr}
    \hline
  $z$ range      & $\delta$ &           $l$ &           $b$ \\
    \hline
    $z \leq 0.2$ &     0.88 & $306^{\circ}$ & $-20^{\circ}$ \\
    $z \leq 0.4$ &     0.52 & $325^{\circ}$ & $ 15^{\circ}$ \\
    $z \leq 0.6$ &     0.20 & $242^{\circ}$ & $-47^{\circ}$ \\
    $z \leq 0.8$ &     0.22 & $310^{\circ}$ & $-30^{\circ}$ \\
    $z \leq 1.0$ &     0.20 & $  1^{\circ}$ & $ -3^{\circ}$ \\
    $z \leq 1.5$ &     0.20 & $352^{\circ}$ & $ -9^{\circ}$ \\
    \hline
\end{tabular}
\end{table*}

\begin{table*}
 \centering
 \caption{Similar to Table \ref{table:union_hemi_tomo}, but for the Constitution data set.}
    \label{table:const_hemi_tomo}
\begin{tabular}{l|rrr}
    \hline
  Redshift range & $\delta$ &           $l$ &           $b$ \\
    \hline
$z \leq 0.6$ &    0.78 &  $ 141^{\circ}$ & $-19^{\circ}$ \\
$z \leq 0.8$ &    0.52 &  $31.6^{\circ}$ & $-15^{\circ}$ \\
$z \leq 1.0$ &    0.51 &  $ 135^{\circ}$ & $ -5^{\circ}$ \\
$z \leq 1.0$ &    0.51 &  $ 135^{\circ}$ & $ -5^{\circ}$ \\
$z \leq 1.55$ &    0.56 &  $ 126^{\circ}$ & $-11^{\circ}$ \\
    \hline
\end{tabular}
\end{table*}

\newpage

\end{document}